\documentclass[1p]{elsarticle}
\makeatletter
\def\ps@pprintTitle{%
 \let\@oddhead\@empty
 \let\@evenhead\@empty
 \def\@oddfoot{\centerline{\thepage}}%
 \let\@evenfoot\@oddfoot}
\makeatother

\usepackage{lineno,hyperref}

\usepackage[labelsep=period,labelfont=bf]{caption} 
\usepackage{amsmath}

\usepackage{hyperref}
\usepackage[hyperref]{xcolor}
\hypersetup{
  colorlinks=true,
}

\begin{document}

\begin{frontmatter}

\title{Wideband acoustic modulation using periodic poroelastic composite structures}


\author[mymainaddress,mymainaddress1]{Hou Qiao}

\author[mymainaddress,mysecondaryaddress]{Zeng He}

\author[mymainaddress,mysecondaryaddress]{Wen Jiang\corref{mycorrespondingauthor}}
\cortext[mycorrespondingauthor]{Corresponding author}
\ead{wjiang@hust.edu.cn}

\author[my4thaddress]{Lin Yang}

\author[my5thaddress]{Weicai Peng}

\address[mymainaddress]{Department of Mechanics, Huazhong University of Science $\&$ Technology, Wuhan, China}

\address[mymainaddress1]{Key Laboratory of Far-shore Wind Power Technology of Zhejiang Province, POWERCHINA HUADONG Engineering Corporation (HDEC), Hangzhou, China}

\address[mysecondaryaddress]{Hubei Key Laboratory for Engineering Structural Analysis and Safety Assessment, Huazhong University of Science $\&$ Technology, Wuhan, China}

\address[my4thaddress]{Department of Engineering Mechanics, Wuhan University of Science and Technology, Wuhan, China}%

\address[my5thaddress]{National Key Laboratory on Ship Vibration and Noise, China Ship Development and Design Center, Wuhan, China}

\begin{abstract}
We proposed an effective acoustic abatement solution comprised of periodic resonators and multi-panel structures with porous lining, which incorporates the wideband capability of porous materials and the low-frequency advantage of locally resonant structures together. Theoretical model and numerical implementation are developed and validated. Two-dimensional poroelastic field expressions are used and resonator forces are incorporated. The results agree well with those reported in the literature and the results obtained from the finite element method. It turns out that sound insulation concerning both the amplitude and tuning bandwidth can be achieved effectively using porous additions and locally-resonant designs. This study presents a promising and practical alternative for wideband acoustic modulation.
\end{abstract}


\end{frontmatter}


\section{Introduction}

Industrial noise, a stubborn disease, has been widely concerned for years; however, its abatement and modulation are very challenging because it exists in a rather wide frequency range and commonly adopted noise reduction methods are only effective for low or medium to high frequency. Therefore, researches on wideband (wide frequency range) noise abatement are exciting and fascinating. Nearly a decade ago, Fuller et al. \cite{Fuller2010NCEJ} found heterogeneous blankets (porous matrix with heterogeneous mass inclusions) that can be adopted as a compact and reliable medium-to-high frequency ($\ge$ 2kHz or 3kHz) noise control solution. Later on, locally tuned resonators were used to enlarge the low-frequency ($\le$ 1kHz to 2kHz) attenuation bandwidth \cite{Xiao2012JSV}; meanwhile, composite structures composed of noise control structures, such as microperforated panels \cite{Liu2017AA,Bucciarelli2019AA} and Helmholtz resonators \cite{Peng2018TJotASoA,Jimenez2017SR}, were utilized to obtain low-frequency wideband performance. Novel porous structures with periodic non-resonant inclusions were also investigated \cite{Groby2008WiRaCM,Groby2011TJotASoA,Nennig2012TJotASoA}, and new acoustic absorption mechanisms were revealed \cite{Groby2008WiRaCM,Groby2011TJotASoA}. Unfortunately, none of those methods can satisfy the requirement of low frequency and medium to high frequency noise reduction at the same time.

Periodic structures, which were introduced by Liu et al. in their seminal work \cite{Liu2000S}, are promising in noise reduction in low-frequency range due to locally resonant mechanism. Unlike in Bragg diffraction cases, locally resonant periodic structures can modulate elastic waves at deep subwavelength. Therefore, to obtain compact and effective low-frequency sound modulation solutions, researchers worldwide focused their interests on composite structures with periodic resonators \cite{Xiao2012JSV,Qiao2019APS}, multi-layer microperforated panel structures \cite{Bucciarelli2019AA,Qiao2019CJoA}, periodic Helmholtz resonators \cite{Fang2006NM} or combined locally resonant structures \cite{Peng2018TJotASoA,Meng2019AA}. Though the results reported in these bibliographies are impressive, locally resonant designs are only effective in a narrow bandwidth, which is an inherent drawback. Therefore, efforts on low-frequency wideband acoustic modulation are prominent. Unfortunately, related results are scarce \cite{Fuller2010NCEJ,Xiao2012JSV,Qiao2019CJoA}, to the authors' knowledge.

On the other hand, porous materials are recognized as wideband noise control solutions. Fruitful theoretical and numerical results on the acoustic properties of porous materials were obtained during the twentieth century \cite{Biot1956TJotASoA,Delany1970AA,Allard1992TJotASoA,Atalla1998TJotASoA,Atalla2001TJotASoA}. As composite structures composed of porous materials, such as layered media \cite{Allard1989JoAP} and multi-layer structures \cite{Bolton1996JSV,Zhou2013JSV}, prevail among noise control applications, a tremendous amount of work about the application of composite structures in sound transmission can be found. As an early attempt, researchers paid their attention to structures made of a porous matrix with inclusions or the so-called 'metaporous materials'. Inspired by the sonic crystal concept, Groby et al. \cite{Groby2008WiRaCM} reported the two bandgaps obtained in a lattice of cylindrical elastic solid scatterers embedded in a rigid-frame porous plate. The influence of equivalent fluid or high-contrast inclusions were reported later \cite{Groby2009}. The combined effects of rigid circular inclusions and irregular rigid backings were also investigated \cite{Groby2011TJotASoA}. Meanwhile, the sound absorption of multilayer porous material with periodic inclusions \cite{Nennig2012TJotASoA} and irregularities \cite{Groby2013} were also reported. Other theoretical or numerical contributions on rigid-frame porous structures with rigid inclusions can also be found \cite{Groby2014,Deckers2016JoCP}. On the other hand, resonant inclusions were introduced into porous matrix (structures) to improve the low-frequency performance. These resonant designs include slotted cylinders \cite{Lagarrigue2013TJotASoA}, Helmholtz resonators \cite{Groby2015TJotASoA,Lagarrigue2016AA}. By incorporating local resonance mechanism to the Bragg’s interference, one can increase the acoustic modulation potential of porous structures.

However, results on poroelastic structures with periodic inclusions or additions are relatively less compared with rigid frame ones. By studying poroelastic heterogeneous acoustic metamaterials, Slagle et al. \cite{Slagle2015} found that the low-frequency acoustic performance of poroelastic structures can be improved by periodic inclusions. Weisser et al. \cite{Weisser2016TJotASoA,Weisser2016} reported that large band absorption could be derived by tuning the properties of the elastic inclusions in the poroelastic domain. Enhanced wave attenuation was also observed by combining standard acoustic foams with locally resonating microstructures \cite{Lewinska2019EJoM-A}. Further investigations are needed to evaluate the potential of novel poroelastic structures. 

The motivation of the present paper is to investigate whether incorporating macroscopic periodic resonators into a poroelastic composite structure is a solution to wideband acoustic modulation. In Section 2, a theoretical model and numerical implementation are prepared to study the sound transmission loss (STL) of the poroelastic composite structures with periodic resonators presented here. In Section 3, the influences of porous additions and resonators are investigated. Section 4 ends with conclusions.

\section{Modeling and validations}

\begin{figure}[!htbp]
     \centering
     \includegraphics[width=1.0\textwidth]{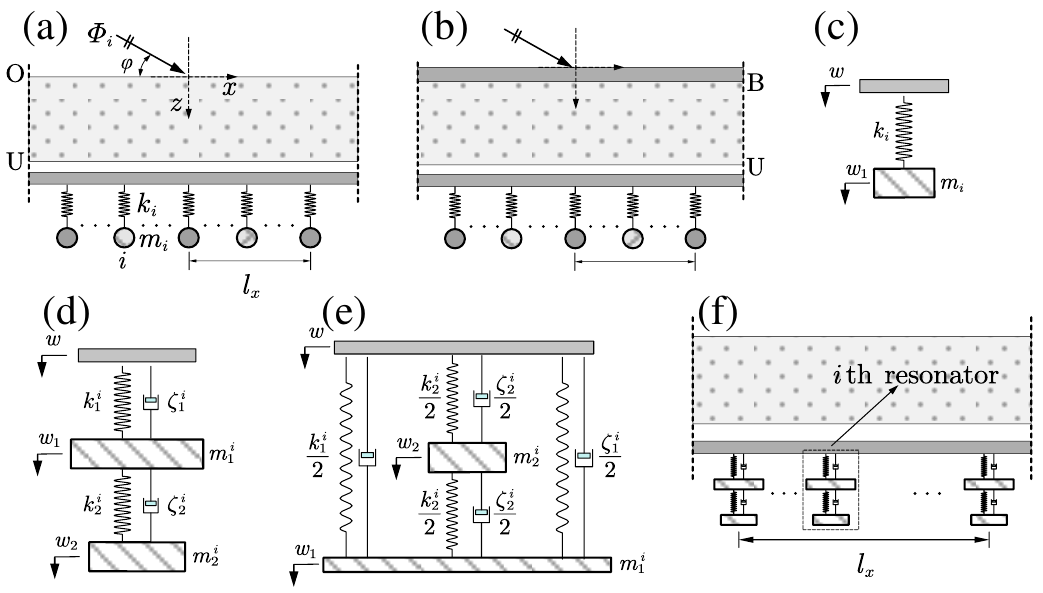}
     \caption{Schematic diagram of the periodic composite structure. (a) A single panel composite structure with porous lining and periodic resonators (OU case, simple resonator). (b) A double panel structure with porous lining and periodic resonators (BU case, simple resonator). (c) A simple resonator. (d) Composite resonator A. (e) Composite resonator B. (f) A single panel composite structure with porous lining and periodic resonators (OU case, composite resonator A).}\label{fig:model2d}
\end{figure}

Two-dimensional periodic poroelastic composite structures, which comprise multi-panel structures with porous lining and distributed resonators, are shown in Fig. \ref{fig:model2d}-a and b. The boundary conditions between the poroelastic domain and an adjacent domain are classified using the notations in Ref. \cite{Bolton1996JSV}, i.e., B for bonded condition, O for open to an infinite air domain, and U for adjacent to but not bonded on a solid domain. The periodic resonators can be simple resonators as shown in Fig. \ref{fig:model2d}-c or composite resonators as shown in Fig. \ref{fig:model2d}-d and e (composite resonator A and B respectively). A composite structure comprised of a single panel with periodic composite resonator A is shown in Fig. \ref{fig:model2d}-f.

A plane wave transmits through the periodic composite structure with velocity potential $\varPhi_i= {\rm e}^{ {\rm j}\omega t-{\rm j} \bf{k} \bf{r}}$, where ${\bf k}=(k_x,k_z)$, ${\bf r}=(x,z)$, ${\rm j}=\sqrt{-1}$. According to Fig. \ref{fig:model2d}-a, $k_x=k \cos\varphi, k_z=k \sin\varphi$, here $k$ is the incident wave number, $\varphi$ is the incident elevation angle. The time dependence ${\rm e}^{{\rm j}\omega t}$ is omitted in the following. Here we make the following assumptions: 
\begin{enumerate}[{A}1.]
\item The composite structure and its adjacent domains are all infinite, therefore no reflections from outside the composite structure are considered.
\item The resonators are periodic along the x-axis with a periodic span $l_x$, and multiple different resonators can be placed evenly in a periodic lattice (span).
\item The composite structure and the resonators are ideally point-connected, therefore the resultant force of a resonator (simple resonator or composite resonator) simplifies to a concentrated force.
\end{enumerate}
A3, however, is not substantial as distributed forces or moments can also be conveniently translated to a concentrated force and/or a moment; it is provided to show the essential idea with elegance here.

The modeling procedures for periodically rib-stiffened composite structure with porous lining \cite{Qiao2019JSV} are adopted and resonator forces are incorporated to develop the present model. The periodicity in the model is treated by truncation and validated by comparing with the finite element results for a degenerate situation.

\subsection{Two-dimensional poroelastic domain with periodic boundary conditions}

When periodic boundary conditions are presented, the poroelastic displacements are assumed to comprise six groups of harmonic components \cite{Qiao2019JSV}. According to the procedures provided in Ref. \cite{Qiao2019JSV}, the displacement ${\bf u}=[{u}_x^s,{u}_z^s,{u}_x^f,{u}_z^f]^T$ in the periodic poroelastic domain is 
\begin{eqnarray}
{\bf u}=\sum_{m}{\rm e}^{-{\rm j} k_x^{m} x }\ {\bf Y}_m{\bf e}_m {\bf C}_m
\end{eqnarray}
where $k_x^{m}$ is the wave number component (along x-axis) of the harmonic components; matrix ${\bf Y}_m$ is the $4\times6$ coefficient matrix of porous field variables (the non-zero elements of ${\bf Y}_m$ are provided in \ref{app:poro_matrices}), which can be obtained using space harmonic series \cite{Qiao2019JSV}; $6\times6$ diagonal matrix ${\bf e}_m$ and $6\times1$ vector ${\bf C}_m$ are
\begin{eqnarray}
{\bf e}_m = diag\left({\rm e}^{{\rm j} k_{1z}^{m} z}, {\rm e}^{-{\rm j} k_{1z}^{m} z}, {\rm e}^{{\rm j} k_{2z}^{m} z}, {\rm e}^{-{\rm j} k_{2z}^{m} z}, {\rm e}^{{\rm j} k_{3z}^{m} z}, {\rm e}^{-{\rm j} k_{3z}^{m} z}\right)\\
{\bf C}_m = [C_1^{m}, C_2^{m}, C_3^{m}, C_4^{m}, C_5^{m}, C_6^{m}]^T
\end{eqnarray}
Here, ${\rm e}^{\pm{\rm j} k_{1z}^{m} z}, {\rm e}^{\pm{\rm j} k_{2z}^{m} z}$, and ${\rm e}^{\pm{\rm j} k_{3z}^{m} z}$ are the wavenumber component (along z-axis) of the harmonic components; $C_1^{m} (i=1,2\ldots 6)$ are the unknown amplitude of the harmonic components (solved later by appropriate boundary conditions). In Ref. \cite{Qiao2019JSV}, a numerical procedure is used to approximate the absence of wave components at normal incidence ($C_5^{m} = C_6^{m}=0$); however, in 2D case, no numerical approximation is needed and the exact closed-form expressions can be obtained.

Once the porous displacement ${\bf u}$ is obtained, the stress components in the poroelastic field is \cite{Biot1956TJotASoA,Allard2009}
\begin{eqnarray}
\sigma_{ij}=2 N e_{ij}+(A e_s + Q e_f)\delta_{ij} \notag\\
s=Q e_s+R e_f\label{eq:biot-stress}
\end{eqnarray}
where $e_s=\partial u_x^s/\partial x+\partial u_z^s/\partial z$, $e_f=\partial u_x^f/\partial x+\partial u_z^f/\partial z$; $N, A, Q$, and $R$ are poroelastic parameters detailed in Ref. \cite{Bolton1996JSV,Allard2009,Qiao2019JSV}, while $e_{ij}$ is
\begin{eqnarray}
e_{ij}=\begin{cases}
\partial u^s_i/\partial x_i,&  i = j \\
\frac{1}{2}\left(\partial u^s_i/\partial x_j+\partial u^s_j/\partial x_i\right),&   i \ne j
\end{cases},\quad
\delta_{ij}=\begin{cases}
1,&  i = j \\
0,&  i \ne j
\end{cases}\notag
\end{eqnarray}
Here $x_i$ and $x_j$ denote $x$ or $z$ coordinate; in other variables, $i, j$ denote $x$ or $z$.

\subsection{In-plane and transverse vibration of the plates}

The thin plate theory is used here as the frequency range discussed is far below the coincidence frequency of the plate; otherwise, the Timoshenko-Mindlin plate theory should be used. Therefore, when in-plane and out-of-plane force or moment are present, the vibration equations of the plates in the periodic poroelastic composite structure here are \cite{Axisa2005}
\begin{eqnarray}
\mathcal{L}_i(u)=f_x ,\quad \mathcal{L}_t(w)=f_z +\frac{\partial \mathcal{M}_y }{\partial x}
\end{eqnarray}
where
\begin{eqnarray}
\mathcal{L}_i(u)=\rho_p h  \frac{\partial^2 u }{\partial t^2}-D_p \frac{\partial}{\partial x}\left( \frac{\partial u }{\partial x} \right),\ 
\mathcal{L}_t(w)=D \nabla^4 w + \rho_p  h  \frac{\partial^2 w }{\partial t^2} \notag
\end{eqnarray}
Here, $\mathcal{L}_i(\cdot)$ and $\mathcal{L}_t(\cdot)$ are the in-plane and out-of-plane vibration operator, $u$ and $w$ are the in-plane and out-of-plane displacements, $f_x$ and $f_z$ are the in-plane and out-of-plane forces, $\mathcal{M}_y$ is the in-plane moment normal to the $x-z$ plane; $\rho_p, h, D_p$, and $D$ are the density, thickness, in-plane stiffness, and the transverse stiffness of the plate.

\subsection{Periodic resonators and their resultant forces}

\begin{figure}[!htbp]
     \centering
     \includegraphics[width=0.90\textwidth]{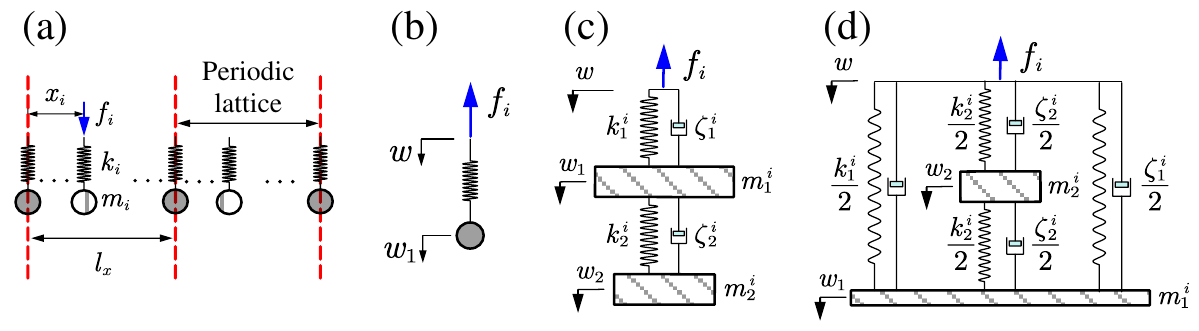}
     \caption{Schematic diagram of the distribution of the spring force. (a) The spring force distribution in a periodic lattice. (b) The displacements and resultant force in a simple resonator. (c-d) The displacements and resultant force in composite resonator A and B respectively.}\label{fig:modelspringforce}
\end{figure}

\begin{figure}[!htbp]
     \centering
     \includegraphics[width=0.90\textwidth]{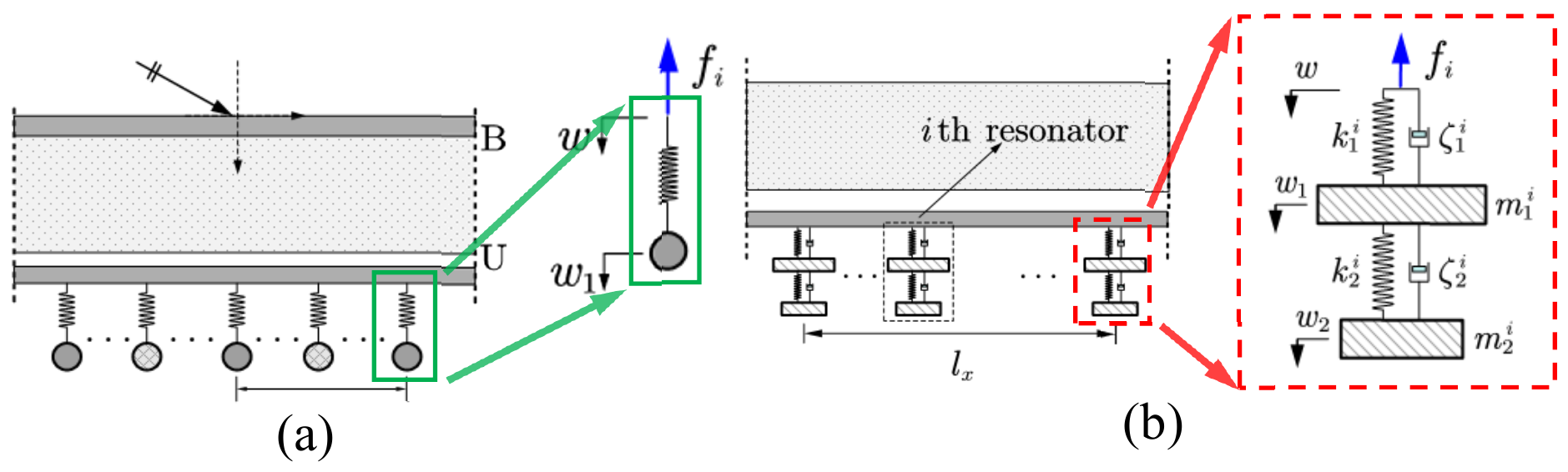}
     \caption{The resonator configurations in a periodic lattice. (a) The case correspond to Fig. \ref{fig:modelspringforce}-b (BU case). (b) The case correspond to Fig. \ref{fig:modelspringforce}-c (OU case).}\label{fig:modelspringforce2}
\end{figure}

The schematic diagram of the placement of spring forces (simple resonators) according to assumptions A2 and A3 is shown in Fig. \ref{fig:modelspringforce}-a; the total number of resonators in a periodic lattice is denoted as $N_s$, and the resonator $i$ ($i=1,2\ldots N_s$) is located at $x_i$ in the periodic span. In Fig. \ref{fig:modelspringforce}-b, \ref{fig:modelspringforce}-c and \ref{fig:modelspringforce}-d, $w$ denotes the displacement of panels where the resonators are attached, as shown in Fig. \ref{fig:modelspringforce2}.  

According to Fig. \ref{fig:modelspringforce}-b, 
\begin{eqnarray}
m_i \ddot{w}_i = - k_i (w_i-w),\
f_i = - k_i (w - w_i) 
\end{eqnarray}
Therefore, the force $f_i$ of a simple resonator $i$ is
\begin{eqnarray}
f_i= \frac{ m_i\omega^2}{1- \omega^2/\omega_i^2}w\ ({\text{without damping}}),\ f_i= \frac{ m_i\omega^2}{1- \omega^2/[\omega_i^2(1+{\rm j}\eta_i)]}w\ ({\text{with damping}})\label{eq:springforce}
\end{eqnarray}
where $\eta_i$ is the spring damping \cite{DenHartog1985}, $w$ is the displacement of the plate to which the resonators are attached; $\omega_i=\sqrt{{k_i}/{m_i}}$ is the natural circular frequency of the resonator $i$, $m_i$ and $k_i$ are the mass and stiffness of the spring.

Meanwhile, according to Fig. \ref{fig:modelspringforce}-c and d, force $f_i$ of a composite resonator $i$ (composite resonator A or B) is
\begin{eqnarray}
f_i = X_3\cdot w
\end{eqnarray}
where ${\bf X}=[X_1,X_2,X_3]^T={\bf H} {\bf F}$, ${\bf H} = ({\bf K}+{\rm j}\omega {\bf C} -\omega^2 {\bf M})^{-1}$; the matrices ${\bf K}, {\bf C}, {\bf M}$, and ${\bf F}$ of composite resonator A or B can all be obtained by their dynamic equations \cite{Qiao2019APS}. Their elements are provided in \ref{app:cr_matrices}.

Therefore, the resultant force $F_{\rm sum}$ of the periodic resonators is
\begin{eqnarray}
F_{\rm sum}=\sum_n \sum_i \beta_i f_i \delta(x-n l_x - x_i) 
\end{eqnarray}
Here, $x_i$ is the position of resonator $i$, $x_i = i a$, $a$ is the distance between two adjacent resonators; integer $n=-\infty,\ldots +\infty$, integer $i=0,1,\ldots N_s$,
\begin{eqnarray}
	\delta(x) =
	\begin{cases}
    1   & \quad x=0\\
    0   & \quad  x \ne 0
  \end{cases}, \quad
  	\beta_i =
	\begin{cases}
    1       & \quad q=1,\ldots (N_s-1)\\
   1/2    & \quad q=0 \text{ or } N_s
  \end{cases}
\end{eqnarray}
$\beta_i$ is the contribution coefficient of a resonator to lattice $n$. 

\subsection{Acoustic domain and boundary conditions}

According to the boundary conditions of the periodic composite structure, there exists one incident acoustic domain, one transmitted acoustic domain, and one or more intermediate acoustic layers between the porous domain and multi-panel structure. The acoustic domains are all assumed linear, and the related velocity potential $\varPhi$ follows
\begin{eqnarray}
\nabla^2 \varPhi - \frac{1}{c^2} \frac{\partial^2 \varPhi}{\partial t^2} = 0 \label{eq:periodic2dwaveeq}
\end{eqnarray}
where $c$ is the corresponding sound velocity.

The boundary conditions between the porous domain and multi-panel structure were detailed before \cite{Bolton1996JSV,Zhou2013JSV,Allard1987}, therefore, not provided here. However, different expressions of boundary conditions should be used for different coordinate systems \cite{Qiao2019JSV}.

\subsection{The harmonic expansions and system equations}

\begin{figure}[!htbp]
     \centering
     \includegraphics[width=0.8\textwidth]{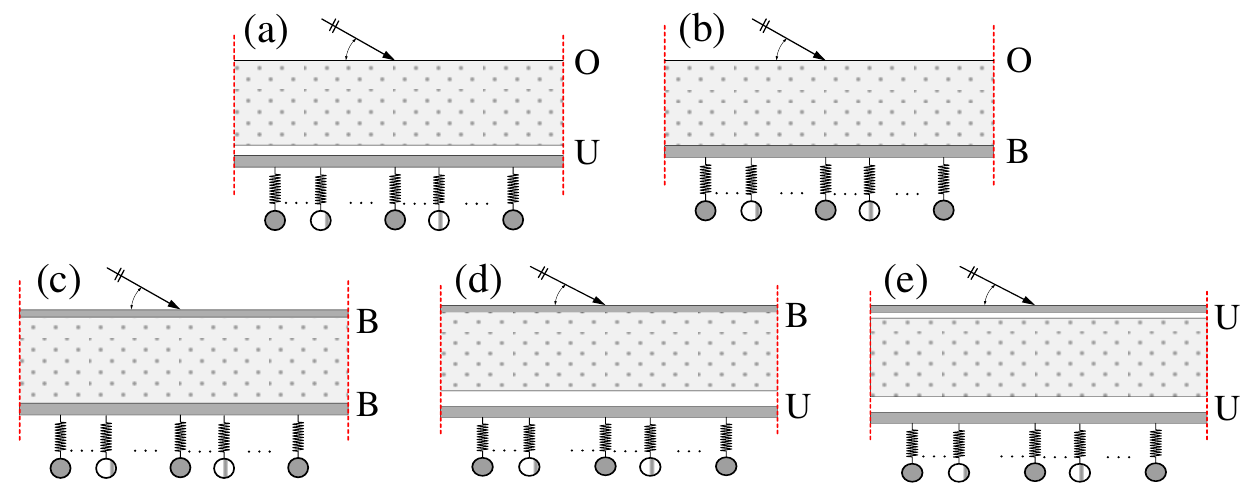}
     \caption{Schematic diagram of different boundary conditions (simple resonators can be replaced by composite resonators): (a-b) single panel case, OU and OB boundary conditions respectively; (c-e) double panel case, BB, BU, and UU boundary conditions respectively. }\label{fig:differentboundaryconditions}
\end{figure}

The periodic composite structure comprises acoustic domains, porous domains, multi-panel structures, and periodic resonators. To show the essential ideas with an elegant formulation, we limit the multi-panel structure to a single panel or a double panel here. The boundary conditions can be OU/OB for the single panel cases (Fig. \ref{fig:differentboundaryconditions}-a and b), and BB/BU/UU for the double panel cases (Fig. \ref{fig:differentboundaryconditions}-c, d, and e). 

In the following, we use the OU case as an example to show the system equations and the solution procedures. In the OU case, the periodic composite structure comprises an incident acoustic domain, a porous domain, an intermediate acoustic domain, a plate, and a transmitted acoustic domain (Fig. \ref{fig:model2d}-a). 

According to the law of refraction and the method of space harmonic series (SHS), the velocity potential of the three acoustic domains can be expressed as \cite{Qiao2019JSV}
\begin{eqnarray}
\varPhi_1 = {\rm e}^{-{\rm j}\left(k_x x+k_z z\right)} + \sum_{m} R_1^m {\rm e}^{-{\rm j}\left(k_x^m x - k_{z,i}^m z\right)} \label{eq:velocitypotential_i} \\
\varPhi_2 = \sum_{m} I_2^m {\rm e}^{-{\rm j}\left(k_x^m x - k_{z,a}^m z\right)} + \sum_{m} R_2^m {\rm e}^{-{\rm j}\left(k_x^m x - k_{z,a}^m z\right)}  \label{eq:velocitypotential_a}\\
\varPhi_3 = \sum_{m} T_3^m {\rm e}^{-{\rm j}\left(k_x^m x - k_{z,t}^m z\right)}  \label{eq:velocitypotential_t}
\end{eqnarray}
where $k_{z,i}^m, k_{z,a}^m$, and $k_{z,t}^m$ are the wave number component (along z-axis) of the harmonic components; $R_1^m, I_2^m, R_2^m$, and $T_3^m$ are the amplitude of the wave harmonics which can be solved by the boundary conditions. $k_{z,i}^m, k_{z,a}^m$, and $k_{z,t}^m$ can be solved by substituting Eq.(\ref{eq:velocitypotential_i}), (\ref{eq:velocitypotential_a}), and (\ref{eq:velocitypotential_t}) into Eq.(\ref{eq:periodic2dwaveeq})
\begin{eqnarray}
k_{z,\times}^m = \sqrt{k^2 - (k_x^m)^2 }, k = \omega/c_{\times} \quad (\times = i, a, {\rm or }\ t)
\end{eqnarray}
Here $c_{\times}$ is the sound velocity in the corresponding domain.

In the thin plate, according to the law of refraction and the method of space harmonic series (SHS), the displacement ${\bf u}=[u,w]^T$ expands to \cite{Qiao2019JSV}
\begin{eqnarray}
{\bf u}=\sum_{m} {\bf U}^m {\rm e}^{-{\rm j} k_x^m x }
\end{eqnarray}
where ${\bf U}^m=[U^m, W^m]^T$ is the unknown component amplitude vector. In OU case, the in-plane displacement $u$ is absent, therefore, only amplitude $W^m$ is included.

In OU case, all the boundary conditions at the interfaces of different domains or at the middle plane are
\begin{eqnarray}
& ({\rm i}) -\epsilon \rho_i \frac{\partial \varPhi_1}{\partial t} - s =0\quad 
({\rm ii}) -(1-\epsilon) \rho_i \frac{\partial \varPhi_1}{\partial t} - \sigma_z=0 
\notag\\
& ({\rm iii}) (1-\epsilon)\frac{\partial u_z^s}{\partial t} + \epsilon \frac{\partial u_z^f}{\partial t} = - \frac{\partial \varPhi_1}{\partial z} \quad 
({\rm iv}) \tau_{zx}=0 
\notag\\
& ({\rm v}) -\epsilon \rho_a \frac{\partial \varPhi_2}{\partial t} - s =0\quad 
({\rm vi}) -(1-\epsilon) \rho_a \frac{\partial \varPhi_2}{\partial t} - \sigma_z=0 
\notag\\
& ({\rm vii}) (1-\epsilon)\frac{\partial u_z^s}{\partial t} + \epsilon \frac{\partial u_z^f}{\partial t} = - \frac{\partial \varPhi_2}{\partial z} \quad 
({\rm viii}) \tau_{zx}=0 
\notag\\
& ({\rm ix}) \frac{\partial w}{\partial t} = -\frac{\partial \varPhi_2}{\partial z} 
\notag\\
& ({\rm x}) D \frac{\partial^4 w}{\partial x^4} + \rho_p h \frac{\partial^2 w}{\partial t^2} = \rho_a \frac{\partial \varPhi_2}{\partial t} - \rho_t \frac{\partial \varPhi_3}{\partial t} +\sum_n \sum_i \beta_i f_i \delta(x-n l_x - x_i) 
\notag\\
& ({\rm xi}) \frac{\partial w}{\partial t} = -\frac{\partial \varPhi_3}{\partial z}\label{eq:systemequation}
\end{eqnarray}
where $\rho_{\times}$ is the density of the related domain; $\epsilon$ is the porosity of the porous media; $\sigma_z, \tau_{zx}$, and $s$ are the normal stress, shear stress and fluid pressure of the solid and fluid phases in the porous media respectively, and they can be obtained using Eq.(\ref{eq:biot-stress}). 

In Eq.(\ref{eq:systemequation}), (i) to (viii) are all applied on the poroelastic-acoustic (inviscid) interface. (i-ii) and (v-vi) correspond to the balance of normal stresses; (iii) and (vii) correspond to the continuity of normal volume velocities; (iv) and (viii) correspond to the shear stress conditions. (ix) and (xi) are the velocity continuity conditions applied on the acoustic-solid (plate) interfaces; (x) is the dynamic equation of the plate applied at its middle plane. The formulation of other boundary conditions can be found in previous papers, for example, Refs. \cite{Bolton1996JSV,Zhou2013JSV,Allard1987}. Therefore, details are not included here.

\subsection{The solution procedures (OU case)}

In Eq.(\ref{eq:systemequation}).(x), double or triple summation occurs; therefore, the resonator force term is rearranged. The cumbersome rearrangement procedures are provided in \ref{app:triplesum}.

Subsequently, according to the orthogonal property below
\begin{eqnarray}
\int_{-l_x/2}^{l_x/2} {\rm e}^{-{\rm j} {k}_{x}^{m}x} {\rm e}^{ {\rm j} {k}_{x}^{p}x} {\rm d}x = 
\begin{cases}
l_x, & m=p \\
0, & m\ne p
\end{cases}\label{eq:orthoproperty}
\end{eqnarray}
where integer $m,p=-\infty,\ldots +\infty$; Eq.(\ref{eq:systemequation}) becomes
\begin{eqnarray}
{\bf A}_m {\bf x}_m=\sum_{n} {\bf B}_n {\bf x}_n + {\bf p} \label{eq:sysmatrix}
\end{eqnarray}
where 
\begin{eqnarray}
{\bf x}_m = [C_1^{m}, C_2^{m}, C_3^{m}, C_4^{m}, C_5^{m}, C_6^{m}, W^m, R_1^m, I_2^m, I_2^m, T_3^m]^T
\end{eqnarray}
Integer $m, n=-\infty,\ldots +\infty$; the non-zero elements of matrices ${\bf A}_m, {\bf B}_n$, and vector ${\bf p}$ are provided in \ref{app:matrixelements}. To obtain a numerical solution, the preceding Eq.(\ref{eq:sysmatrix}) is rearranged using 
\begin{eqnarray}
\tilde{\bf x}=[\cdots,\ {\bf x}_{m-1}^T,\ {\bf x}_{m}^T,\ {\bf x}_{m+1}^T,\ \cdots]^T \label{eq:Xrearranged}
\end{eqnarray}
\begin{eqnarray}
\tilde{\bf A}=
\begin{bmatrix}
\ddots &   &  & &      \\
 & {\bf A}_{m-1} & & & \\
 &  &{\bf A}_m& & \\
 &  & &{\bf A}_{m+1} & \\
  &  & &  & \ddots 
\end{bmatrix} \label{eq:Arearranged}
\end{eqnarray}
\begin{eqnarray}
\tilde{\bf B}=
\begin{bmatrix}
 &   & \vdots &  &  \\
\cdots & {\bf B}_{n-1} &{\bf B}_{n} &{\bf B}_{n+1} &\cdots  \\
\cdots & {\bf B}_{n-1} &{\bf B}_{n} &{\bf B}_{n+1} &\cdots \\
\cdots & {\bf B}_{n-1} &{\bf B}_{n} &{\bf B}_{n+1} &\cdots \\
 &  & \vdots &  & 
\end{bmatrix} \label{eq:Brearranged}
\end{eqnarray}
\begin{eqnarray}
\tilde{\bf p}=[\cdots,\ {\bf 0}^T,\ {\bf p}^T,\ {\bf 0}^T,\ \cdots]^T \label{eq:prearranged}
\end{eqnarray}
Here, the blank blocks of $\tilde{\bf A}$ are all zeros, while $\tilde{\bf B}$ is a full block matrix; ${\bf 0}^T$ is a zero $1\times 14$ vector. Subsequently, Eq.(\ref{eq:sysmatrix}) becomes 
\begin{equation}
(\tilde{\bf A}-\tilde{\bf B}) \tilde{\bf x}=\tilde{\bf p} \label{eq:finalmatrix}
\end{equation}
As the dimension of matrices $\tilde{\bf A}, \tilde{\bf B}$, and vectors $\tilde{\bf x}$, $\tilde{\bf p}$ are all infinite, Eq.(\ref{eq:finalmatrix}) is an infinite matrix equation and can be solved by truncation. When $m$ is truncated to $[-\hat{m}, \hat{m}]$, $\tilde{\bf A}$ and $\tilde{\bf B}$ are truncated to $14\hat{M}\times 14\hat{M}$ matrices $\bar{\bf A}$ and $\bar{\bf B}$, vectors $\tilde{\bf x}$ and $\tilde{\bf p}$ as $14\hat{M}\times 1$ vectors $\bar{\bf x}$ and $\bar{\bf p}$, $\hat{M}=2\hat{m}+1$; therefore, 
\begin{equation}
 \bar{\bf x}=(\bar{\bf A}-\bar{\bf B})^{-1} \bar{\bf p} \label{eq:matrixsolution}
\end{equation}
Subsequently, the unknown amplitude of the harmonic components are determined.

The sound transmission coefficient $\tau$ of the composite structure is \cite{Qiao2019JSV,Legault2009JSV}
\begin{eqnarray}
\tau(\varphi) = \frac{\rho_t}{\rho_i} \frac{\sum_{m}|T_3^m|^2 {\rm Re}(k_{z,t}^m) }{{\rm Re}(k_z)}
\end{eqnarray}
Here ${\rm Re}(\cdot)$ is the real operator for a complex variable. When sound wave transmit through the composite structure at $\varphi\in [\varphi_{\rm lim}, \pi/2]$, where $\varphi_{\rm lim}$ is the minimum elevation angle (that sound can transmit), the STL of the preceding composite structure is defined as
\begin{eqnarray}
{\rm STL}= 10 \log (1/\bar{\tau}),\ \bar{\tau}=\frac{\int_{\varphi_{\rm lim}}^{\pi/2} \tau(\varphi) \sin\varphi \cos\varphi\ {\rm d}\varphi}{\int_{\varphi_{\rm lim}}^{\pi/2} \sin\varphi \cos\varphi\ {\rm d}\varphi} 
\end{eqnarray}
When the unknown amplitudes are obtained, STL can be solved by numerical integration subsequently.

\subsection{Validation}

The convergence characteristics of the periodic poroelastic composite structure are complicated \cite{Qiao2019JSV}, therefore, a convergence check step, which is confirmed as reliable and accurate \cite{Qiao2019JSV,Legault2009JSV,Xin2010JotMaPoS}, is performed before any further computations to determine the appropriate $\hat{m}$. The convergence criteria is chosen as $\Delta {\rm STL}=0.1$dB at the maximum computation frequency, i.e., when the variation in STL by changing $\hat{m}$ (add or minus by one) is less than 0.1dB, it is considered as converged.

To validate the present periodic composite model and show its applicability in a wide frequency range, we compared the results obtained from the present model with the results reported in the Refs. \cite{Xiao2012JSV,Qiao2019APS} for low-frequency and high-frequency situations and those obtained from FEM. The parameters in Ref. \cite{Xiao2012JSV,Bolton1996JSV,Qiao2019APS} are adopted (as in Tab. \ref{tab:parameters} below); however, the FEM setups are not detailed here as they are identical to those in Ref. \cite{Qiao2019JSV} except that it is 2D here. The details of Ref. \cite{Xiao2012JSV} and Ref. \cite{Qiao2019APS} are briefly revisited here to facilitate readers' understanding.

In Ref. \cite{Qiao2019APS}, a periodic poroelastic composite structure comprised of three parts, i.e. poroelastic domain, elastic domain (thin plate), and periodic resonators (simple single-degree-of-freedom resonators or composite two-degree-of-freedom resonators) is proposed. Though Biot theory is used for the poroelastic domain in Ref. \cite{Qiao2019APS}, the resonator-plate coupling system is assumed to be an anisotropic plate with an equivalent dynamic density (i.e. effective medium), which is only valid in the low-frequency range. Therefore, the applicable frequency range, and the modeling method in Ref. \cite{Qiao2019APS} are different from those here, though the model configurations in Ref. \cite{Qiao2019APS} are identical to the OU and OB cases here (Fig.\ref{fig:model2d}-a and b).

Meanwhile, Ref. \cite{Xiao2012JSV} was chosen to validate the models here for the reason that the metamaterial plates in Ref. \cite{Xiao2012JSV} can be represented by the OU and OB cases here (Fig.\ref{fig:model2d}-a and b) by choosing the porosity as 1, and the air gap thickness as zero in the OU case, or just choosing the porosity as 1 in the OB case with the corresponding boundary conditions.

\subsubsection{Low-frequency validations}

In the low-frequency range, comparisons with the multiple identical simple resonator (resonance frequency $f_r$=300Hz) or composite resonator (resonance frequency $f_1^0$=300Hz) results in Ref. \cite{Xiao2012JSV,Qiao2019APS} are made.

\begin{figure}[!htbp]
     \centering
     \includegraphics[width=0.85\textwidth]{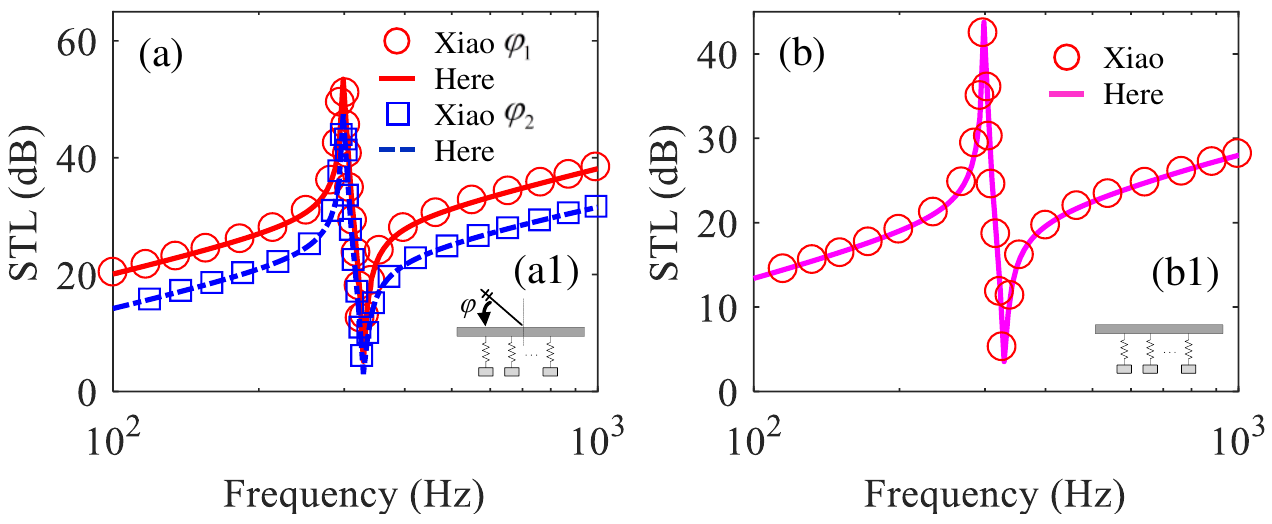}
     \caption{Comparison of predictions (lines) versus Ref.  \cite{Xiao2012JSV} (symbols) (a) oblique incidence case $\varphi_1=\pi/2$ or $\varphi_2=\pi/6$ and (b) random incidence case, (a1) and (b1) are the schematic diagram respectively. The frequency range studied is [100Hz, 1kHz]. }\label{fig:valxiaoobliqueanddiffuself}
\end{figure}

\begin{figure}[!htbp]
     \centering
     \includegraphics[width=.95\textwidth]{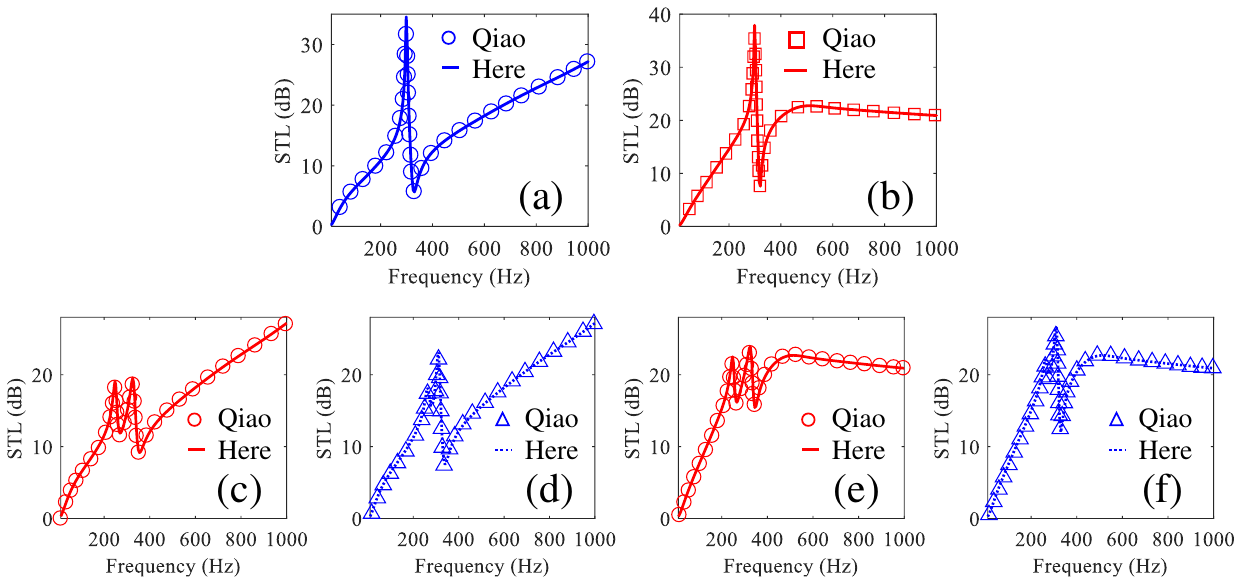}
     \caption{Comparison of predictions (lines) versus Ref.  \cite{Qiao2019APS} (symbols): (a-b) multiple simple resonators, OU and OB respectively; (c-d) multiple composite resonator A or B, OU case; (e-f) multiple composite resonator A or B, OB case. The frequency range studied is [10Hz, 1kHz]. }\label{fig:valqiao1peqpmodel}
\end{figure}

Either the oblique incidence case results (Fig. \ref{fig:valxiaoobliqueanddiffuself}-a) or the random incidence case results (Fig. \ref{fig:valxiaoobliqueanddiffuself}-b and Fig. \ref{fig:valqiao1peqpmodel}-a to f) all show excellent consistency.

\begin{figure}[!htbp]
     \centering
     \includegraphics[width=0.95\textwidth]{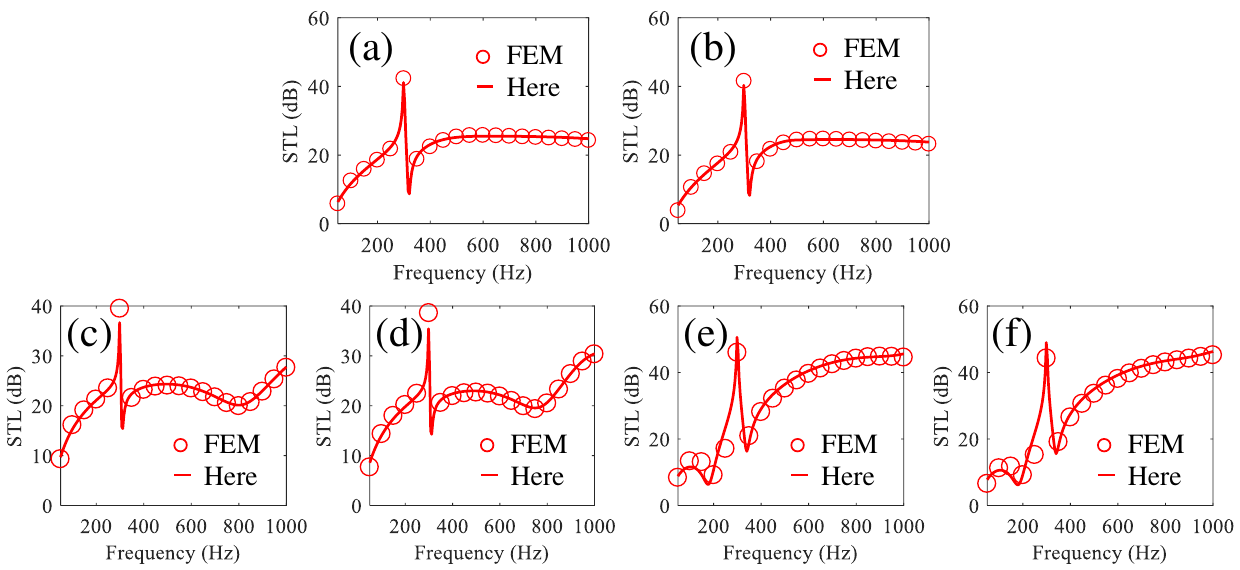}
     \caption{Comparison of predictions (lines) versus FEM results (symbols): (a-b) OB case, $\varphi=\pi/2,\pi/3$ respectively; (c-d) BB case, $\varphi=\pi/2,\pi/3$ respectively; (e-f) UU case, $\varphi=\pi/4,\pi/6$ respectively. The frequency range studied is [10Hz, 1kHz]. }\label{fig:valfemresultslf}
\end{figure}

Some of the oblique incidence FEM results are given in Fig. \ref{fig:valfemresultslf} (OB, BB, and UU cases). The overall consistency is satisfactory. The results in Fig. \ref{fig:valxiaoobliqueanddiffuself}, Fig. \ref{fig:valqiao1peqpmodel}, and Fig. \ref{fig:valfemresultslf} confirmed the validity here in the low-frequency range.

\subsubsection{High-frequency validations}

In the high-frequency range, result comparisons versus Ref. \cite{Xiao2012JSV} and FEM are presented (the resonance frequency $f_r$ or $f_1^0$ is 3kHz if not specified).

\begin{figure}[!htbp]
     \centering
     \includegraphics[width=0.75\textwidth]{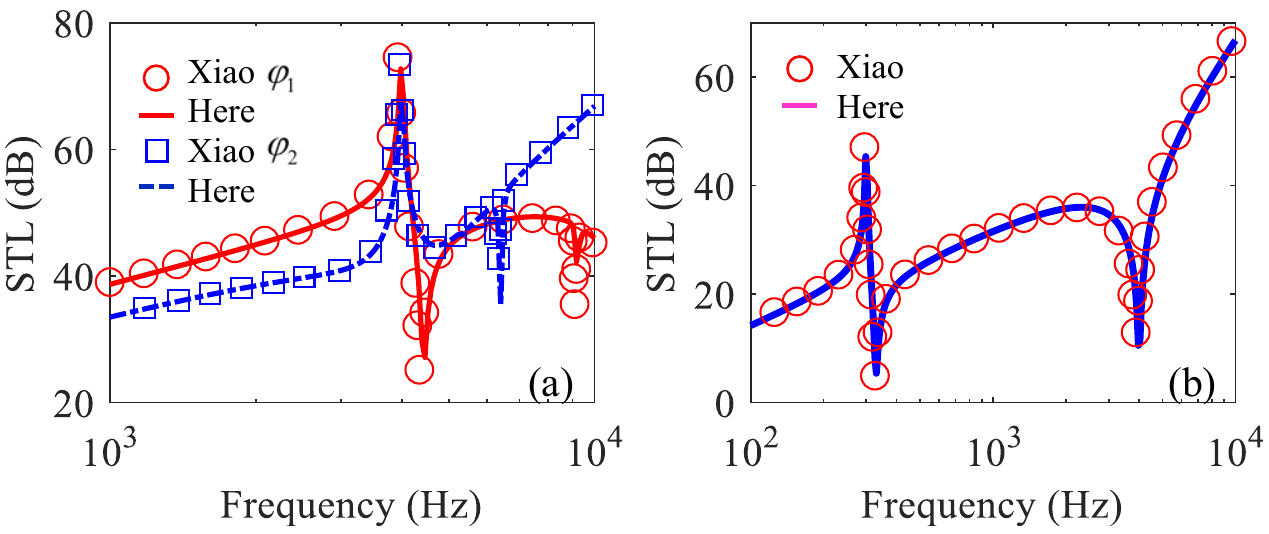}
     \caption{Comparison of predictions (lines) versus Ref. \cite{Xiao2012JSV} (symbols) (a) oblique incidence case $\varphi_1=\pi/3$ or $\varphi_2=\pi/6$, $f_r$=4kHz; (b) random incidence case, $f_r$=300Hz. The frequency range studied is [1kHz, 10kHz] and [100Hz, 10kHz] respectively. }\label{fig:valxiaoobliqueanddiffusehf}
\end{figure}

Comparisons versus Ref. \cite{Xiao2012JSV} show satisfactory overall consistency (Fig. \ref{fig:valxiaoobliqueanddiffusehf}).

\begin{figure}[!h]
     \centering
     \includegraphics[width=0.95\textwidth]{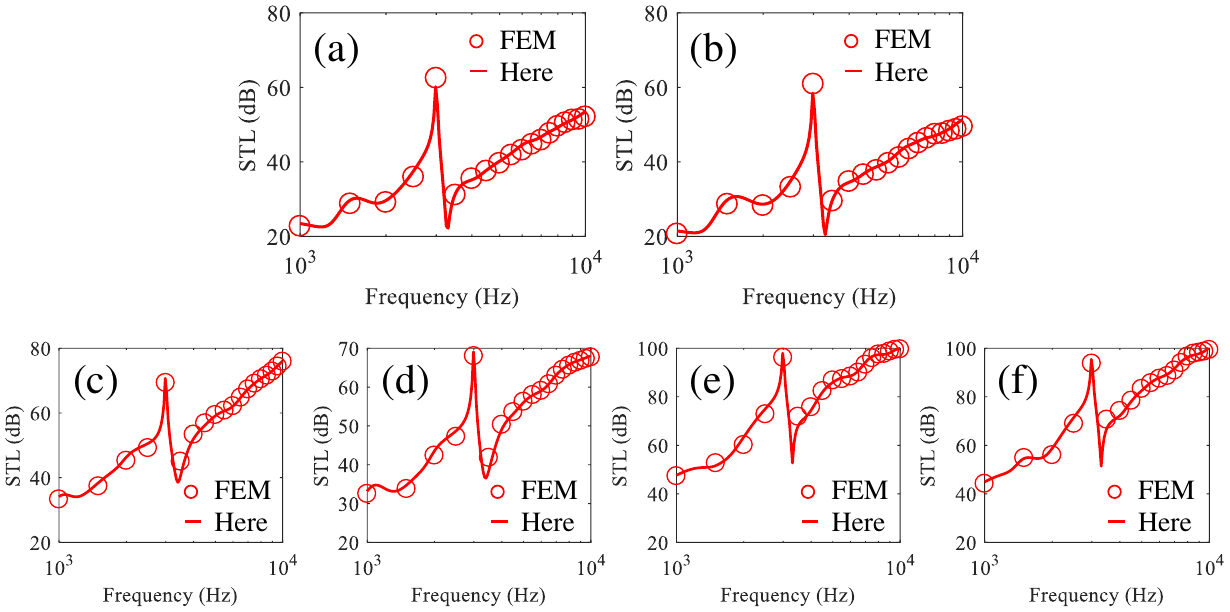}
     \caption{Comparison of predictions (lines) versus FEM results (symbols): (a-b) OB case, $\varphi=\pi/4,\pi/6$ respectively; (c-d) BB case, $\varphi=\pi/4,\pi/6$ respectively; (e-f) BU case, $\varphi=\pi/2,\pi/3$ respectively. The frequency range studied is [1kHz, 10kHz]. }\label{fig:valfemresultshf}
\end{figure}

Some oblique incidence result comparisons versus FEM results are shown in Fig. \ref{fig:valfemresultshf} (OB, BB, and BU cases). The results agree well with those obtained from FEM. Fig. \ref{fig:valxiaoobliqueanddiffusehf} and Fig. \ref{fig:valfemresultshf} confirmed the validity in the high-frequency range.

\section{Results and discussions}
Default parameter values adopted in the following study are listed in Tab. \ref{tab:parameters} unless otherwise specified. The STL is calculated in 1/24 octave bands between 10Hz and 10kHz using Simpson's Rule; the integration domain is split into 90 subdivisions. In 2D case, different minimum elevation angle $\varphi_{\rm lim}$ are used \cite{Bolton1996JSV,Xiao2012JSV,Legault2009JSV}, either by experiment data or theoretical prediction. To be consistent with previous studies, we choose $\varphi_{\rm lim}=0$ for the single panel case \cite{Xiao2012JSV,Legault2009JSV} and $\varphi_{\rm lim}=\pi/10$ for the double panel case \cite{Bolton1996JSV}. As STL is sensitive to the total mass \cite{Hambric2016}, here the mass ratio $\gamma$ of the resonators, i.e. the percentage of the resonator mass in the total mass, is kept constant as $\gamma=0.2$ \cite{Xiao2012JSV,Qiao2019APS} in the following.

\begin{table}[!h]
\caption{Model parameters: air gap thickness $h_{a}$ = 2 mm for the OU case; $h_{a}$ = 14 mm for the BU case; $h_{a1}$ = 2mm, $h_{a2}$ = 6 mm for the two air gaps in the UU case; $l_x=h_p, (h_p+h_a), h_p, (h_p+h_a)$, and $(h_p + h_{a1}+ h_{a2})$ for the OB, OU, BB, BU, and UU cases, respectively; the gap properties $\rho_g=\rho_i, c_g=c_i$ and the transmitted side media properties $\rho_t=\rho_i, c_t=c_i$}\label{tab:parameters}
\begin{tabular}{lll}
\hline
Parameters & Descriptions & Value\\
\hline
Acoustic media\\
$\rho_i$ & density (incident side) & 1.205 kg/${\rm m}^{3}$\\
$c_i$ & sound velocity (incident side) & 343 m/${\rm s} $\\
\hline
Panels\\
$h_1$ & panel thickness (incident/single panel)  & 1.270 mm\\
$h_2$ & panel thickness (transmitted) & 0.762 mm\\
$\rho_p$ & density of face panels & 2700 kg/${\rm m}^{3}$\\
$E_p$ & Young's modulus of face panels & 70$\times 10^9$Pa\\
$\nu_p$ & Poisson's ratio of face panels & 0.33\\
\hline
Porous media\\
$\rho_s$ & bulk density of solid phase & 30 kg/${\rm m}^{3}$\\
$\rho_f$ & density of fluid phase & 1.205 kg/${\rm m}^{3}$\\
$E_s$ & Young's modulus (solid phase) & 8$\times 10^5$Pa\\
$\nu_s$ & Poisson's ratio (solid phase) & 0.4\\
$\eta_s$ & loss factor (solid phase) & 0.265\\
$\epsilon$ & the porosity & 0.9\\
$\epsilon'$ & the tortuosity & 7.8\\
$\sigma$ & flow resistivity & 2.5$\times 10^4$ MKS\ Rayls/${\rm m} $\\
$h_p$ & thickness of porous core & 27 mm\\
\hline
\end{tabular}
\end{table}


\subsection{Influence of porous additions on the STL}

\begin{figure}[!htbp]
     \centering
     \includegraphics[width=0.75\textwidth]{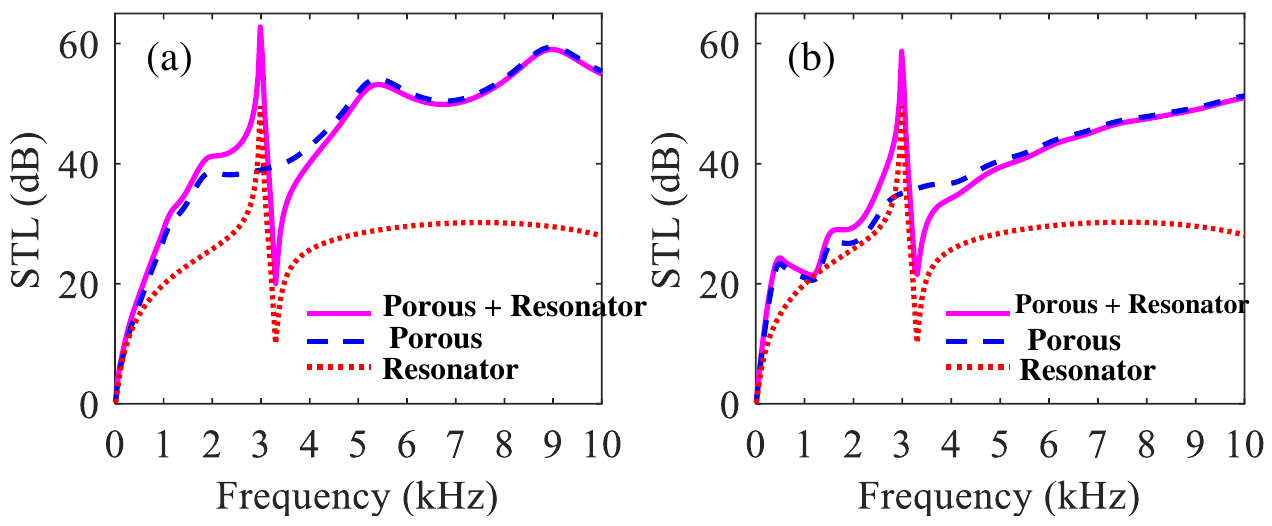}
     \caption{Influence of porous additions on the STL (a) OU case and (b) OB case. }\label{fig:porousaddingresultsr1}
\end{figure}

To show the influence of porous additions, here we used multiple identical simple resonators in a periodic span, where $m_i$=27g, $\eta_i$=0.01. 
Fig. \ref{fig:porousaddingresultsr1} (OU and OB cases) shows the comparisons between the STL of the periodic composite structures (porous + resonator, $f_r$=3kHz), multi-panel structures (porous), and plate with periodic simple resonators (metamaterial plate, $f_r$=3kHz).

From Fig. \ref{fig:porousaddingresultsr1}, we can see that porous addition can improve the STL of the periodic composite structure away from the resonance frequencies, and ease the STL decrease around the local resonance. As porous media has wideband attenuation capability, its wideband improvement is conceivable; meanwhile, it can attenuate the sound transmission even though resonance is prominent around the resonance frequencies. In summary, the STL of the periodic composite structure can be considered as the superposition of individual contributions from porous additions and resonators together.

\subsection{Influence of identical simple resonators on the STL}

\begin{figure}[!htbp]
     \centering
     \includegraphics[width=0.8\textwidth]{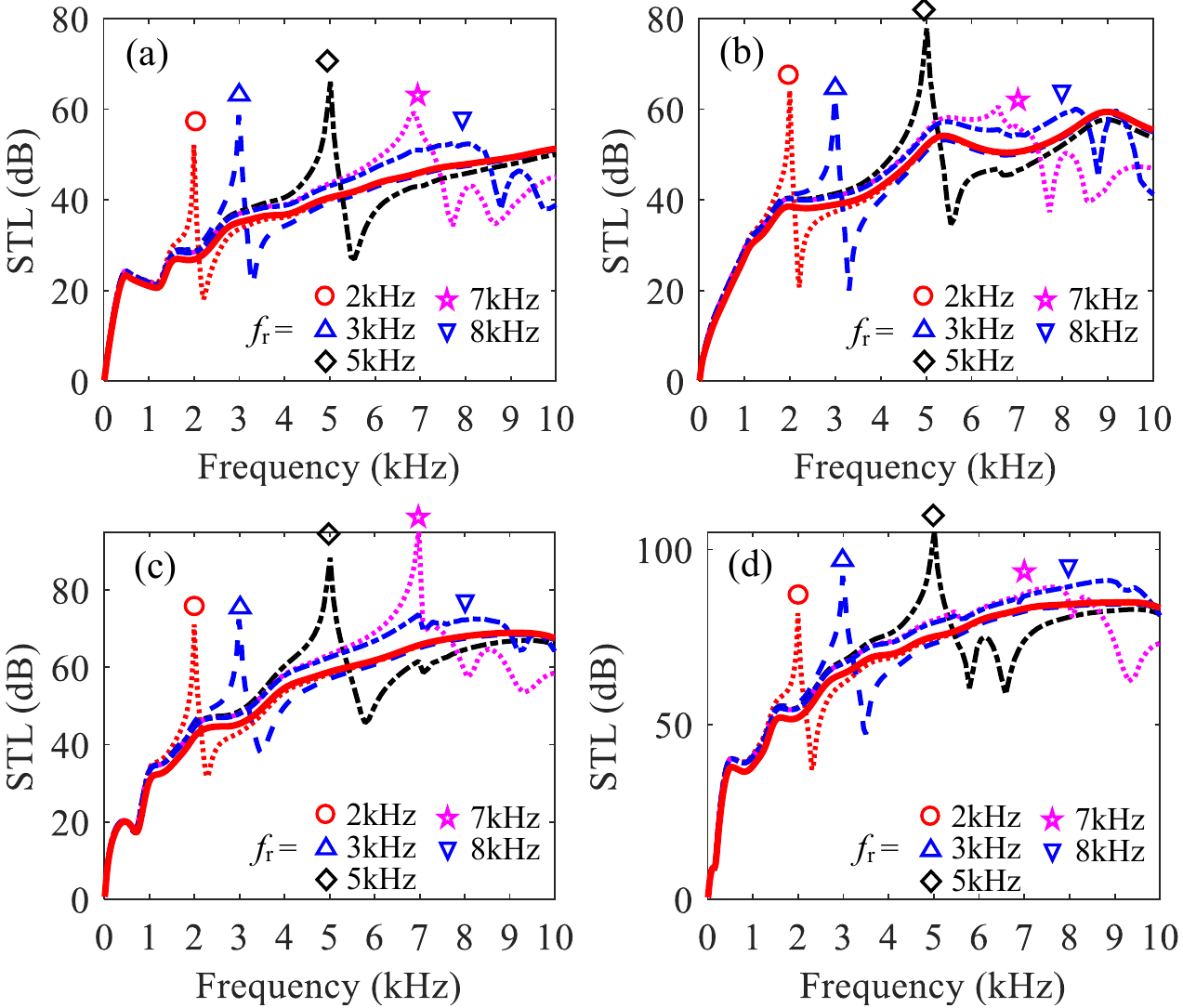}
     \caption{Influence of identical simple resonators on the STL (a) OB case, (b) OU case, (c) BB case, and (d) BU case. Different lines (markers) correspond to different resonance frequency cases; the solid lines show the results without resonators. $f_r$ is the resonance frequency of the resonators.}\label{fig:multiplesimplertrsresultsr1}
\end{figure}

Fig. \ref{fig:multiplesimplertrsresultsr1} shows the results of the periodic composite structure here with multiple identical simple resonators (different resonance frequencies respectively) under different boundary conditions (OB, OU, BB, and BU cases). Here $m_i$=27g, $\eta_i$=0.01. 

Notable STL improvement occurs around the resonance frequencies (a local crest), with a decrease around the subsequent anti-resonance frequencies (a local trough). The overall STL recovers to the cases without resonators (the solid lines in Fig. \ref{fig:multiplesimplertrsresultsr1}) away from these resonance frequencies. However, while the STL increase around the resonance frequency $f_r$=2kHz, 3kHz, and 5kHz is evident, which resembles the performance of resonators \cite{Xiao2012JSV} and the STL is then dominated by the resonators, the improvement around higher resonance frequencies weakens (Fig. \ref{fig:multiplesimplertrsresultsr1}-a and c $f_r$=8kHz, Fig. \ref{fig:multiplesimplertrsresultsr1}-b and d $f_r$=7, 8kHz); the STL shows multiple troughs subsequently. The extra troughs are due to the coupling between resonators and the corresponding composite structures. Quantitative analyses on the dispersion relations are in progress.

\subsection{Influence of identical composite resonators on the STL}

Composite (multiple-degree-of-freedom) resonators can provide better vibration or wave attenuation \cite{Peng2015IJoMS}. Therefore, two composite resonator cases (Fig. \ref{fig:model2d}-d and e) are used here to evaluate the sound insulation performance. In composite resonator $i$, we denote $m_2^i=r\cdot m_1^i$, $k_2^i=s\cdot  k_1^i$, $\zeta_2^i=t \cdot  \zeta_1^i$, and the damping ratio $\eta_n^i=\zeta_n^i/2 m_n^i \omega_n^i$, $\omega_n^i=\sqrt{k_n^i/m_n^i}$ ($n$=1, 2 for the primary and secondary resonators respectively). If no damping is present (i.e. $\zeta_1^i=\zeta_2^i=0$), the resonance frequencies of composite resonator A are \cite{Peng2015IJoMS,Qiao2019APS}
\begin{eqnarray}
f_{1,2} =\frac{\omega_1^i}{2\pi} \sqrt{\frac{r+s+r s\pm \sqrt{(r+s+r s)^2-4r s}}{2r}} \label{eq:compraomega12}
\end{eqnarray}
The resonance frequencies of composite resonator B are \cite{Qiao2019APS}
\begin{eqnarray}
f_{1,2} =\frac{\omega_1^i}{2\pi} \sqrt{\frac{2r+2s+r s\pm \sqrt{(2r-2s+r s)^2+4r s^2}}{4r}} \label{eq:comprbomega12}
\end{eqnarray}

\begin{figure}[!h]
     \centering
     \includegraphics[width=0.8\textwidth]{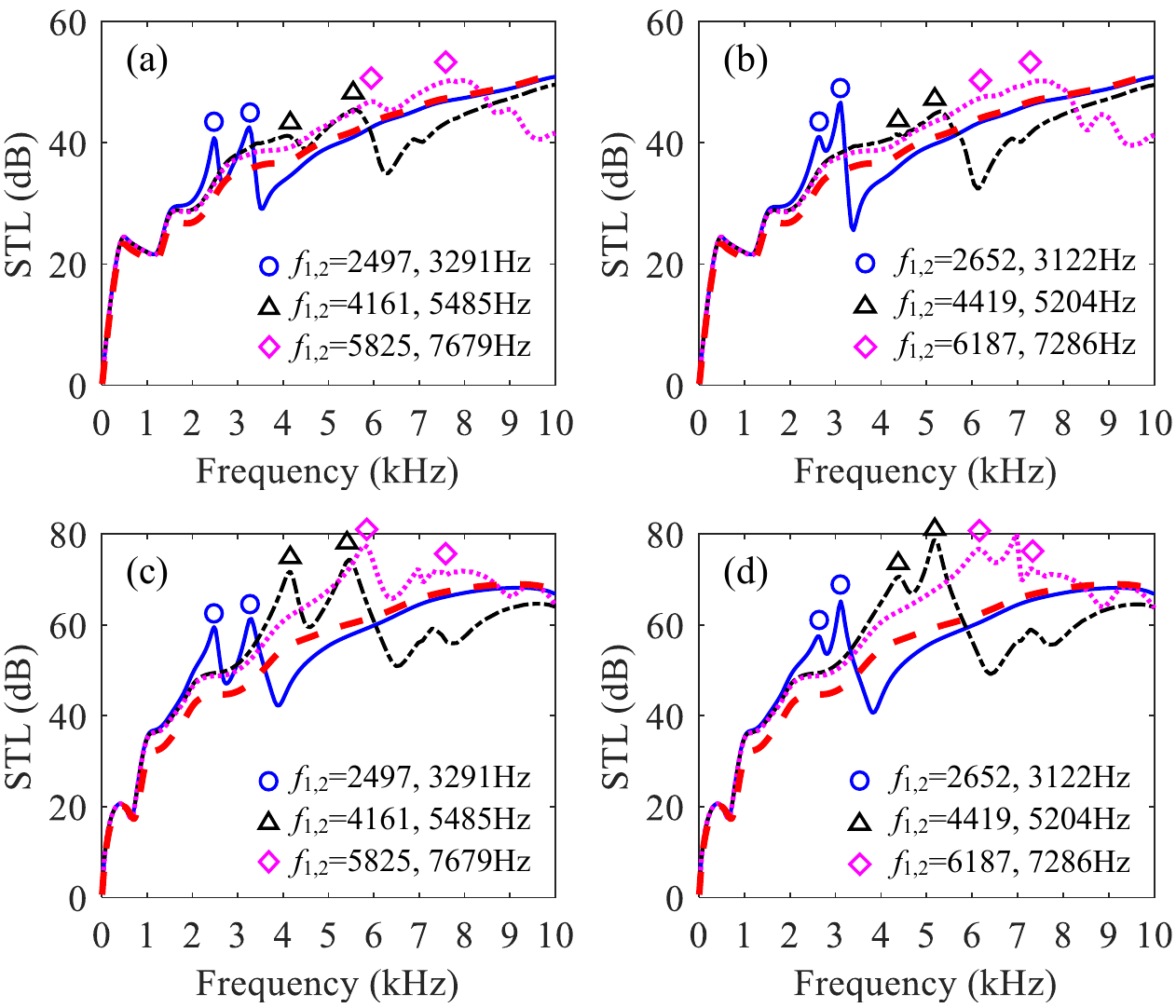}
     \caption{Influence of identical composite resonators on the STL (a-b) OB case (composite A and B respectively) and (c-d) BB case (composite A and B respectively). Different lines (markers) correspond to different resonance frequency cases; the thick dashed lines (unmarked) show the results without resonators. $f_{1,2}$ are the resonance frequencies of composite resonators.}\label{fig:multiplecomprtrsresultsr1}
\end{figure}

Fig. \ref{fig:multiplecomprtrsresultsr1} shows the results of the periodic composite structures here with multiple identical composite resonators (A or B, with different resonance frequencies) under different boundary conditions (OB and BB cases). Parameter values in Tab. \ref{tab:m1crparameters} are used. 

Notable STL improvements occur around the resonance frequencies $f_{1,2}$ (local crests), with decreases around the subsequent anti-resonance frequencies (local troughs). The overall STL recovers to the cases without resonators (the thick dashed lines in Fig. \ref{fig:multiplecomprtrsresultsr1}) away from the resonance frequencies. However, as $f_{1,2}$ increase, the improvement around resonance frequencies weakens; additional STL troughs emerge subsequently. The extra troughs are analogous to the cases with multiple identical simple resonators. One thing that should be noted is that, while the STL improvement around higher resonance frequencies weakens, the higher resonance frequencies contribute to wider sound insulation modulation bandwidth versus cases without resonators.

\begin{table}[!htbp]
	\begin{center}
		\caption{Parameters of the composite resonators. }
		\label{tab:m1crparameters}
		\begin{tabular}{ccccccc}
			\hline
			Parameters & $\gamma$ & $m_1^i$ & $r$ & $s$ & $\eta_1^i$ & $\eta_2^i$\\
			\hline
			Values & 0.2 & 30 g & 0.0750 & 0.0625 & 0.01 & 0.05 \\
			\hline
		\end{tabular}
	\end{center}
\end{table}

\subsection{Influence of multiple simple resonators on the STL}

\begin{figure}[!htbp]
     \centering
     \includegraphics[width=0.8\textwidth]{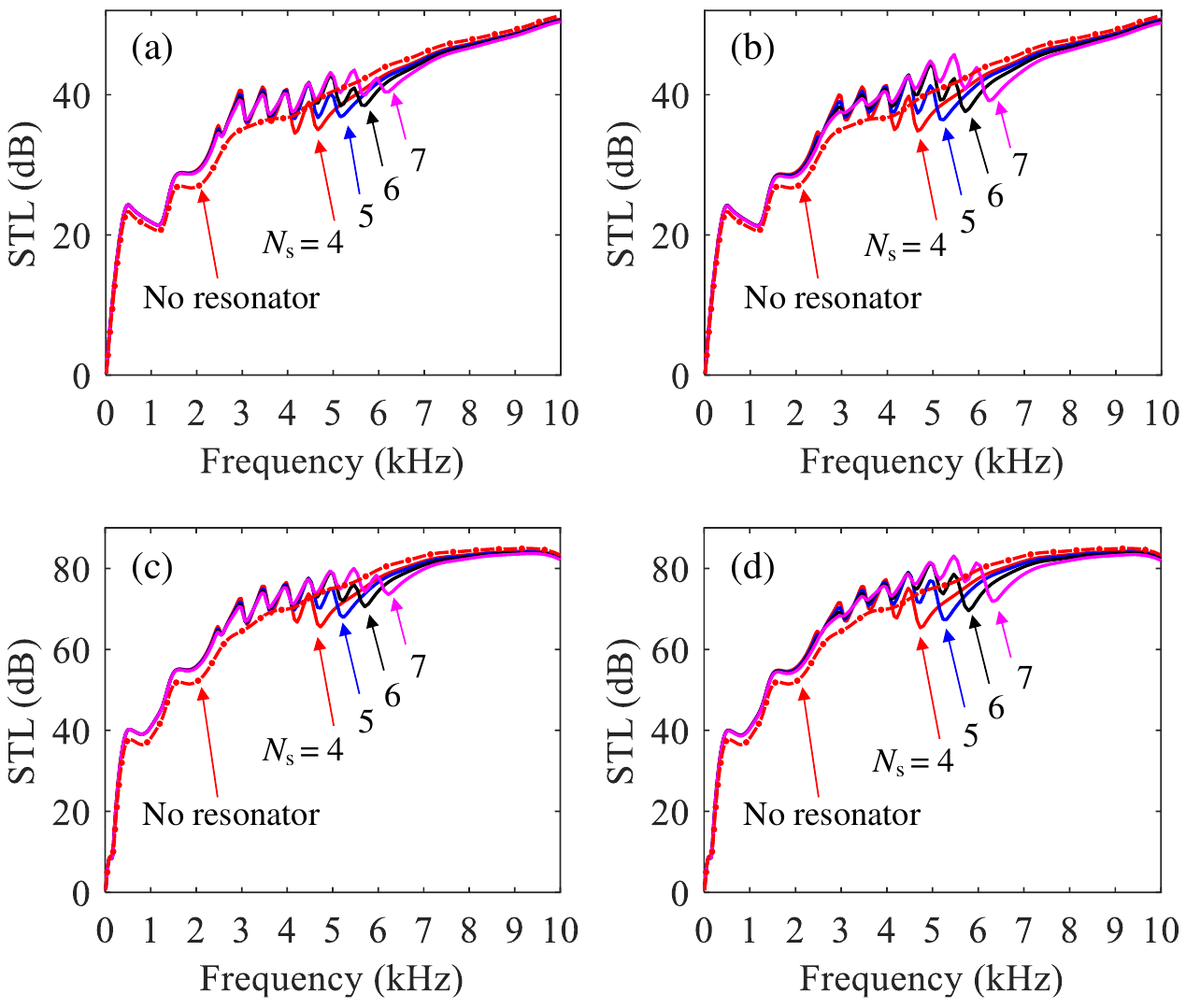}
     \caption{Influence of multiple simple resonators on the STL (a-b) OB case (case I and II respectively) and (c-d) BU case (case I and II respectively). $N_s$ is the total number of resonators in a periodic lattice.}\label{fig:multipledifferentsimprtrsresultsr1}
\end{figure}

Two multiple simple resonator cases are investigated here using the parameters in Tab. \ref{tab:mmsrparameters}. The STL results of the periodic composite structures are given in Fig. \ref{fig:multipledifferentsimprtrsresultsr1} (OB and BU boundary conditions, case I and II respectively). 

\begin{table}[!h]
	\begin{center}
		\caption{Parameters of multiple simple resonator cases I and II, the mass ratio $\gamma$=0.2; for resonator $i$, the mass $m_i$ and resonance frequency $f_r^i$ are $m_i=m_0+ (i-1)\Delta m$, $f_r^i=3+0.5(i-1)$(kHz), $i=1,\ldots N_s$, $N_s$=4,\ldots 7.}
		\label{tab:mmsrparameters}
		\begin{tabular}{cllll}
			\hline
			Case & Description & $m_0$ & $\Delta m$ & $\eta_i$\\
			\hline
			I  & Constant $m_i$ & $m_{\rm sum}/N_s$ & 0 & 0.05\\
			\hline
			II  & Increasing $m_i$ & $m_{\rm sum}/N_s - (N_s-1) \Delta m /2$ & $0.04 m_{\rm sum}$ & 0.05\\
			\hline
		\end{tabular}
	\end{center}
\end{table}

The STL can be tuned around the resonance frequencies of the simple resonators. When the resonance frequencies are in the low-frequency range, the sound insulation of the periodic composite structure can be improved; however, as the resonance frequencies increase, the sound insulation deteriorates and becomes not better than the cases without resonators. As the resonator number $N_s$ increases, the STL amplitude and its modulation bandwidth can be tuned in either case I or II. Furthermore, while cases I and II are analogous, their STL are slightly different. A comparison is made in the following.

The local STL crests and troughs are caused by the resonance and anti-resonance of the periodic resonators. The extremum amplitudes and bandwidths are determined by the damping and resonance frequencies of the resonators. If multiple simple resonators are properly selected, the STL amplitudes and the bandwidths can be tuned, therefore, the sound insulation performance here is modulated.

\subsection{Influence of multiple composite resonators on the STL}

\begin{figure}[!htbp]
     \centering
     \includegraphics[width=0.8\textwidth]{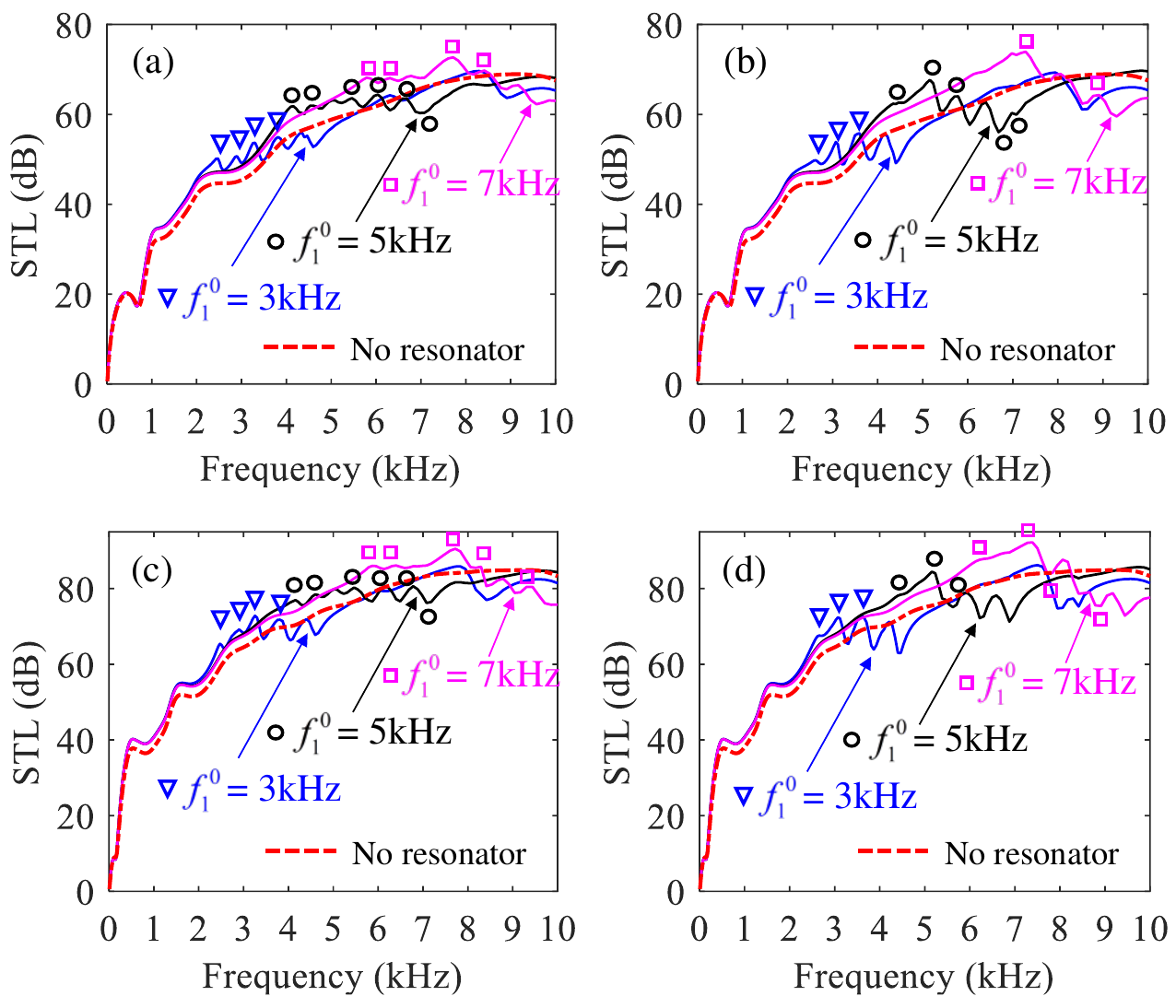}
     \caption{Influence of multiple composite resonators on the STL (a-b) BB case (different composite resonator A and B cases respectively) and (c-d) BU case (different composite resonator A and B cases respectively). The markers show the resonance frequencies (partial, clearly identifiable ones) in the corresponding cases, and the arrows differentiate resonator cases. $f_1^0$ is the first resonance frequency of resonator 1 in a periodic lattice. }\label{fig:mmcrresultsr1}
\end{figure}

Composite resonators with different resonance frequencies are arranged in a single lattice using parameters in Tab. \ref{tab:mmcrparameters}; in resonator $i$, the resonance frequency of the primary mass is chosen as $f_1^i=f_1^0+500(i-1)$(Hz), $i=1,\ldots N_s$. The results are given in Fig. \ref{fig:mmcrresultsr1} (BB and BU boundary conditions, multiple composite resonator A or B cases respectively).

\begin{table}[!h]
	\begin{center}
		\caption{Parameters of the composite resonators}
		\label{tab:mmcrparameters}
		\begin{tabular}{lcccccc}
			\hline
			Parameters & $\gamma$ & $N_s$ & $r$ & $s$ & $\eta_1^i$ & $\eta_2^i$ \\
			\hline
			Values & 0.2 & 4 & 0.0750 & 0.0625 & 0.01 & 0.05 \\
			\hline
		\end{tabular}
	\end{center}
\end{table}

Apart from the resonance frequencies obtained by Eq.(\ref{eq:compraomega12}) and (\ref{eq:comprbomega12}), when damping is included, the characteristic frequencies of composite resonators (A or B) can be determined by a quartic equation
\begin{eqnarray}
A_4 \omega^4 + A_3 \omega^3 + A_2 \omega^2 + A_1 \omega + A_0 = 0
\end{eqnarray}
For composite resonator A, the coefficients are $A_4=r$, $A_3=-2 {\rm j} (r + t + r t) \eta_1 \omega_1$, $A_2=- (r+s+r s+4t \eta_1^2)\omega_1^2$, $A_1= 2 {\rm j} (s + t) \eta_1 \omega_1^3$, and $A_0=s \omega_1^4$; as to composite resonator B, $A_4= r$, $A_3=-{\rm j} (2 r +2 t + r t) \eta_1 \omega_1$, $A_2=- \left[ r+s+\frac{1}{2} r s+(4t + t^2) \eta_1^2 \right] \omega_1^2$, $A_1= {\rm j} (2 s +2 t + s t) \eta_1 \omega_1^3$, and $A_0= \frac{1}{4}\left(4s + s^2\right) \omega_1^4$. The characteristic frequencies are determined by solving these quartic equations.

The STL can be improved around the resonance frequencies in the low-frequency range; however, as the resonance frequencies increase, the STL deteriorates and becomes not better than the cases without resonators (Fig. \ref{fig:mmcrresultsr1}). This is analogous to the cases of simple resonators. Furthermore, the influence of the lower resonance frequencies is noteworthy compared to the higher ones; besides, the influence of the latter tends to vanish, and distinct STL troughs can even be seen due to complex coupling between the resonators and the structures. Quantitative analyses are in progress at the moment.

\subsection{Comparison between different resonator cases}

\begin{figure}[!h]
     \centering
     \includegraphics[width=0.8\textwidth]{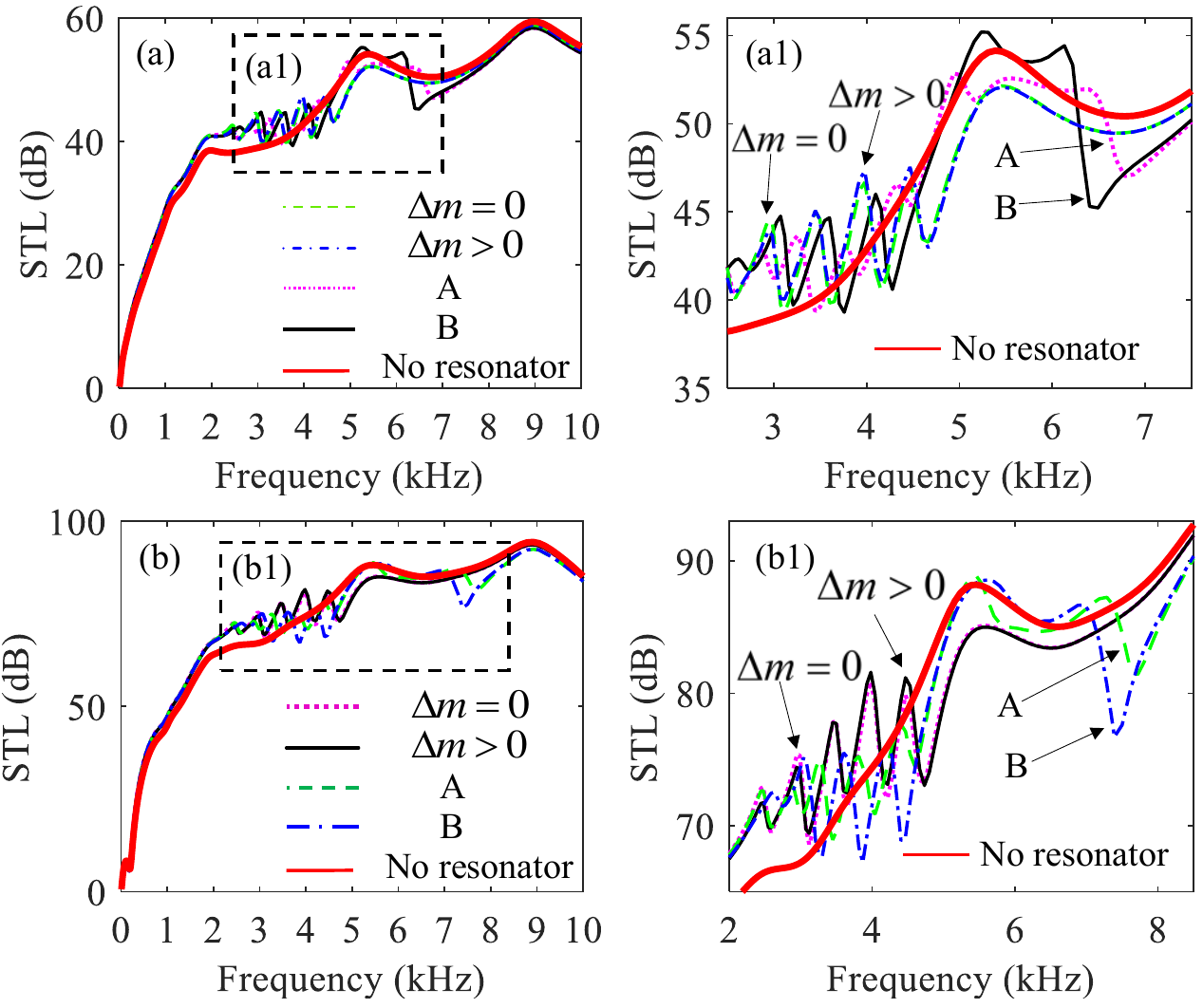}
     \caption{Comparison of the four resonator cases (a-a1) OU case and (b-b1) UU case; (a1) and (b1) shows the partial highlighted view of (a) and (b). $\Delta m=0$ and $\Delta m>0$ correspond to the multiple simple resonator cases I and II respectively, while A and B correspond to multiple composite resonator A and B cases respectively. }\label{fig:compallresultsr1}
\end{figure}

To show the differences between the above four resonator cases, we choose 
$N_s$=4, $\gamma$=0.2 (other parameters as those in Tab. \ref{tab:mmsrparameters} and \ref{tab:mmcrparameters}), and the resonance frequencies ($f_r^i$ or $f_1^i$) as $3000 + 500(i-1)$(Hz), $i=1,\ldots N_s$. All the results are given in Fig. \ref{fig:compallresultsr1}. 

The bandwidths of the two simple resonator cases are almost coincident as the resonance frequencies are identical, while the STL magnitudes are different because of different mass distribution; however, the STL differences are tiny as the mass differences of the resonators are not significant. 
Compared with simple resonator cases, the composite resonator A and B cases all show wider working bandwidth, without a significant decrease in STL magnitudes. 
However, the bandwidth or STL magnitude is slightly different in the two composite resonator cases. In Fig. \ref{fig:compallresultsr1}, composite resonator B cases have slightly larger STL magnitudes, but with narrower working bandwidth, and more drastic STL decreases beyond these resonance frequencies. 

In a word, multiple simple resonator cases apply to simple STL modulation solutions, as both the bandwidths and magnitudes can be tuned as desired; however, if wider bandwidth is anticipated and a tradeoff in magnitudes are acceptable, composite resonator cases can be the right alternatives.

\section{Conclusions}\label{stn:conclusion}
A poroelastic multi-panel structure with porous lining and periodic resonators embedded is presented in this paper to achieve wideband acoustic modulation. The wideband capability of porous materials and low-frequency advantage of periodic structures can be utilized and incorporated together in that structure. The STL of the present periodic composite structure is caused by both porous additions and resonators; therefore, both the amplitude and tuning bandwidth of STL can be adjusted with appropriate resonator arrangements. When the resonance frequencies of the resonators are low, notable STL improvements occur nearby, with a decrease around the subsequent anti-resonance frequencies (a local trough); however, as the resonance frequencies increase, the STL modulation nearby weakens. Compared to simple resonators, adopting composite resonators can obtain wider bandwidths while a compromise in the magnitude of STL is inevitable. Though the results are preliminary, the synergistic effect of locally resonant design and poroelastic composite structures is quite impressive, which is likely to provide a promissing wideband acoustic modulation solution.

\section*{Acknowledgements}
This work is supported by the National Natural Science Foundation of China (NSFC) No.11572137. Part of this work is conducted within the doctoral research of Qiao, whose doctoral thesis is going to be published in Chinese.

\appendix

\section{The non-zero elements in matrices of composite resonators}\label{app:cr_matrices}

The non-zero elements in matrices ${\bf K}, {\bf C}, {\bf M}$, and vector ${\bf F}$ of composite resonator A are
\begin{eqnarray}
& K(1,1) = k_2^i,\ K(1,2) = -k_2^i,\ K(2,1) = -k_1^i -k_2^i \notag
\\
& K(2,2) = k_2^i,\ K(3,1) = k_1^i, \ K(3,3) = -1 \notag
\\
& C(1,1) = \zeta_2^i,\ C(1,2) =- \zeta_2^i,\ C(2,1) =- \zeta_1^i - \zeta_2^i \notag
\\
& C(2,2) = \zeta_2^i,\ C(3,1) = \zeta_1^i \notag
\\
& M(1,2) = -m_2^i,\ M(2,1) = -m_1^i \notag
\\
& F(2) = -k_1^i -{\rm j} \omega \zeta_1^i,\ F(3) = k_1^i +{\rm j} \omega \zeta_1^i
\end{eqnarray}

The non-zero elements in matrices ${\bf K}, {\bf C}, {\bf M}$, and vector ${\bf F}$ of composite resonator B are
\begin{eqnarray}
& K(1,1) = -k_1^i - \frac{k_2^i}{2},\ K(1,2) = \frac{k_2^i}{2},\ K(2,1) =\frac{k_2^i}{2} \notag
\\
& K(2,2) = - k_2^i,\ K(3,1) = k_1^i,\ K(3,2)=\frac{k_2^i}{2},\ K(3,3) = -1 \notag
\\
& C(1,1) = -\zeta_1^i - \frac{\zeta_2^i}{2},\ C(1,2) = \frac{\zeta_2^i}{2},\ C(2,1) = \frac{\zeta_2^i}{2} \notag
\\
& C(2,2) =- \zeta_2^i,\ C(3,1) = \zeta_1^i,\ C(3,2)=\frac{\zeta_2^i}{2} \notag
\\
& M(1,1) = -m_1^i,\ M(2,2) = -m_2^i \notag
\\
& F(1) = -k_1^i -{\rm j} \omega \zeta_1^i,\ F(2) =  -\frac{k_2^i}{2} -{\rm j} \omega \frac{\zeta_2^i}{2} \notag
\\
& F(3) = k_1^i +\frac{k_2^i}{2} +{\rm j} \omega  \zeta_1^i +{\rm j} \omega \frac{\zeta_2^i}{2}
\end{eqnarray}

\section{The coefficient matrices of poroelastic variables}\label{app:poro_matrices}
The non-zero elements of coefficient matrix ${\bf Y}_m$ are
\begin{eqnarray}
&Y_m(1,1)=Y_m(1,2)=\frac{{\rm j} k_x^m }{k_1^2},\ Y_m(1,3)=Y_m(1,4)=\frac{{\rm j} k_x^m }{k_2^2}\notag
\\
&Y_m(1,5)=\frac{{\rm j} k_{3z}^m }{k_3^2},\ Y_m(1,6)=-\frac{{\rm j} k_{3z}^m }{k_3^2}\notag
\\
&Y_m(2,1)=\frac{{\rm j} k_{1z}^m }{k_1^2},\ Y_m(2,2)=-Y_m(2,1),\ Y_m(2,3)=\frac{{\rm j} k_{2z}^m }{k_2^2},\ Y_m(2,4)=-Y_m(2,3)\notag
\\
&Y_m(2,5)=Y_m(2,6)=-\frac{{\rm j} k_x^m }{k_3^2}\notag
\\
&Y_m(3,1)=Y_m(3,2)=b_1 \frac{{\rm j} k_x^m }{k_1^2},\ Y_m(3,3)=Y_m(3,4)=b_2 \frac{{\rm j} k_x^m }{k_2^2}\notag
\\
&Y_m(3,5)=g \frac{{\rm j} k_{3z}^m }{k_3^2},\ Y_m(3,6)=-g \frac{{\rm j} k_{3z}^m }{k_3^2}\notag
\\
&Y_m(4,1)=b_1 \frac{{\rm j} k_{1z}^m }{k_1^2},\ Y_m(4,2)=-Y_m(4,1),\ Y_m(4,3)=b_2 \frac{{\rm j} k_{2z}^m }{k_2^2},\ Y_m(4,4)=-Y_m(4,3)\notag
\\
&Y_m(4,5)=Y_m(4,6)=-g \frac{{\rm j} k_x^m }{k_3^2}\notag
\end{eqnarray}
Here, $k_1, k_2, k_3, b_1, b_2$, and $g$ were described in detail in Ref. \cite{Bolton1996JSV,Qiao2019JSV}.

\section{The summation in Eq.(\ref{eq:systemequation}).(x)}\label{app:triplesum}

According to Eq.(\ref{eq:springforce}) and the Poisson summation formula \cite{Qiao2019JSV}
\begin{eqnarray}
\sum_m \delta(x - m l_x-x_i) = \frac{1}{l_x} \sum_m {\rm e}^{{\rm j}2\pi m (x-x_i)/l_x}\label{eq:poissonsum}
\end{eqnarray}
The summation
\[
	S=\sum_{i=0}^{N_s} \sum_m \beta_i F_i \delta(x - m l_x-x_i) 
\]
becomes
\begin{eqnarray}
S = \frac{1}{l_x} \sum_{i=0}^{N_s}\frac{ m_i\omega^2 \beta_i}{1- \omega^2/[\omega_i^2(1+{\rm j}\eta_i)]} \sum_m \sum_n {\rm e}^{{\rm j}2\pi m (x- x_i)/l_x}  W_n {\rm e}^{-{\rm j} k_x^n x} \label{eq:tripleexp1}
\end{eqnarray}

Noting the following identity (derivation can be found in \ref{app:derivdoublesum}),
\begin{eqnarray}
\sum_{n=-\infty}^{+\infty} \sum_{m=-\infty}^{+\infty} W_m {\rm e}^{-{\rm j} {k}_{x}^{m}x} {\rm e}^{{\rm j} 2\pi n (x-x_i)/l_x}=\sum_{n=-\infty}^{+\infty} W_n {\rm e}^{{\rm j} 2\pi n (-x_i)/l_x} \sum_{m=-\infty}^{+\infty} {\rm e}^{-{\rm j} {k}_{x}^{m}x} {\rm e}^{{\rm j} 2\pi (-m) (-x_i)/l_x} \label{eq:identify}
\end{eqnarray}
Eq.(\ref{eq:tripleexp1}) can be written as
\begin{eqnarray}
S = \frac{1}{l_x} \sum_{i=0}^{N_s}\frac{ m_i\omega^2 \beta_i}{1- \omega^2/[\omega_i^2(1+{\rm j}\eta_i)]} \sum_{n} W_n {\rm e}^{{\rm j} 2\pi n (-x_i)/l_x} \sum_{m} {\rm e}^{-{\rm j} {k}_{x}^{m}x} {\rm e}^{{\rm j} 2\pi (-m) (-x_i)/l_x} \label{eq:tripleexp2}
\end{eqnarray}
According to the orthogonality described by Eq.(\ref{eq:orthoproperty}), Eq.(\ref{eq:tripleexp2}) can be written as
\begin{eqnarray}
 \frac{1}{l_x} \int_{-l_x/2}^{l_x/2} S {\rm e}^{{\rm j} {k}_{x}^{p}x} {\rm d}x= \frac{1}{l_x} \sum_{n} W_n {\rm e}^{{\rm j} 2\pi (n-p) (-x_i)/l_x} \sum_{i=0}^{N_s}\frac{ m_i\omega^2 \beta_i}{1- \omega^2/[\omega_i^2(1+{\rm j}\eta_i)]} \label{eq:triplereduced}
\end{eqnarray}
Then Eq.(\ref{eq:sysmatrix}) can be obtained by rearranging Eq.(\ref{eq:systemequation}) and utilizing Eq.(\ref{eq:triplereduced}) at the same time.
 
\section{Derivation of the double summation identity }\label{app:derivdoublesum}

As ${\rm exp}(-{\rm j} {k}_{x}^{m}x)\cdot {\rm exp}({\rm j} 2\pi n x/l_x) = {\rm exp}[-{\rm j} {k}_{x}^{(m-n)}x]$, the left-hand side of Eq.(\ref{eq:identify}) becomes
\begin{eqnarray}
\sum_{n=-\infty}^{+\infty}\left( \cdots +W_{-1}\ {\rm e}^{-{\rm j} {k}_{x}^{-1-n}x}+W_{0}\ {\rm e}^{-{\rm j} {k}_{x}^{-n}x}+W_{1}\ {\rm e}^{-{\rm j} {k}_{x}^{1-n}x}+\cdots \right) {\rm e}^{{\rm j} 2\pi n (-x_i)/l_x}\label{eq:identityp1}
\end{eqnarray}

As $n = -\infty, \ldots \infty $, if the index $n$ loops over $(n-1)$ to $(n+1)$, Eq.(\ref{eq:identityp1}) becomes
\begin{eqnarray}
\begin{split}
\Big\{\cdots&+\left( \cdots +W_{-1} \ {\rm e}^{-{\rm j} {k}_{x}^{-n}x}+W_{0} \ {\rm e}^{-{\rm j} {k}_{x}^{-n+1}x}+W_{1} \ {\rm e}^{-{\rm j} {k}_{x}^{-n+2}x}+\cdots \right){\rm e}^{{\rm j} 2\pi (n-1) (-x_i)/l_x}+\\
&+\left( \cdots +W_{-1} \ {\rm e}^{-{\rm j} {k}_{x}^{-n-1}x}+W_{0} \ {\rm e}^{-{\rm j} {k}_{x}^{-n}x}+W_{1} \ {\rm e}^{-{\rm j} {k}_{x}^{-n+1}x}+\cdots \right){\rm e}^{{\rm j} 2\pi (n+0) (-x_i)/l_x}+\\
&+\left( \cdots +W_{-1} \ {\rm e}^{-{\rm j} {k}_{x}^{-n-2}x}+W_{0} \ {\rm e}^{-{\rm j} {k}_{x}^{-n-1}x}+W_{1} \ {\rm e}^{-{\rm j} {k}_{x}^{-n}x}+\cdots \right){\rm e}^{{\rm j} 2\pi (n+1) (-x_i)/l_x}+\cdots\Big\}
\end{split}\label{eq:identityp2}
\end{eqnarray}

Collecting exponential terms in Eq.(\ref{eq:identityp2}), one can obtain
\begin{eqnarray}
\begin{split}
\Big\{\cdots
&+{\rm e}^{-{\rm j} {k}_{x}^{-n-1}x}{\rm e}^{-(-n-1)*}\left(\cdots+W_{-1}{\rm e}^{-1*}+W_{0}{\rm e}^{0*}+W_{1}{\rm e}^{1*}+\cdots \right)+\cdots\\
&+{\rm e}^{-{\rm j} {k}_{x}^{-n}x}{\rm e}^{-(-n)*}\left(\cdots+W_{-1}{\rm e}^{-1*}+W_{0}{\rm e}^{0*}+W_{1}{\rm e}^{1*}+\cdots \right)+\cdots\\
&+{\rm e}^{-{\rm j} {k}_{x}^{-n+1}x}{\rm e}^{-(-n+1)*}\left(\cdots+W_{-1}{\rm e}^{-1*}+W_{0}{\rm e}^{0*}+W_{1}{\rm e}^{1*}+\cdots \right)+\cdots
\Big\}
\end{split}\label{eq:identityp3}
\end{eqnarray}
where ${\rm e}^{p*}={\rm e}^{{\rm j} 2\pi p (-x_i)/l_x}, p\in Z$. Subsequently,  Eq.(\ref{eq:identityp3}) becomes
\begin{eqnarray}
\sum_{n} \sum_{m} W_m {\rm e}^{-{\rm j} {k}_{x}^{m}x} {\rm e}^{{\rm j} 2\pi n (x-x_i)/l_x}=\sum_{n} W_n {\rm e}^{n*} \sum_{m} {\rm e}^{-{\rm j} {k}_{x}^{m}x} {\rm e}^{-m*}
\end{eqnarray}
which is
\begin{eqnarray}
\sum_{n} \sum_{m} W_m {\rm e}^{-{\rm j} {k}_{x}^{m}x} {\rm e}^{{\rm j} 2\pi n (x-x_i)/l_x}=\sum_{n} W_n {\rm e}^{{\rm j} 2\pi n (-x_i)/l_x} \sum_{m} {\rm e}^{-{\rm j} {k}_{x}^{m}x} {\rm e}^{{\rm j} 2\pi (-m) (-x_i)/l_x}
\end{eqnarray}

\section{The non-zero elements of the matrices in Eq.(\ref{eq:sysmatrix})}\label{app:matrixelements}

Denoting $k_x^m, k_{z,i}^m, k_{z,a}^m, k_{z,a}^m, k_{1z}^m, k_{2z}^m$, and $k_{3z}^m$ as $\alpha_m, \gamma_{i,m}, \gamma_{a,m}, \gamma_{t,m}, \gamma_{1,m}, \gamma_{2,m}$, and $\gamma_{3,m}$ respectively; $L_1=h_p+h_a$, $L_2=h_p+h_a+\frac{h_1}{2}$, $L_3=h_p+h_a+{h_1}$, the non-zero elements of matrix ${\bf A}_m$ are

\begin{eqnarray}
& A_m(1,1)=-Q_0-b_1 R_0,\ A_m(1,2)=A_m(1,1),\ A_m(1,3)=-Q_0-b_2 R_0 \notag
\\
&A_m(1,4)=A_m(1,3),\ A_m(1,8)=-{\rm j}\rho_i \epsilon \omega  \notag 
\\
& A_m(2,1)=-A_0 - b_1 Q_0 -2 N_0 \frac{\gamma_{1,m}^2}{k_1^2},\ A_m(2,2)=A_m(2,1) \notag
\\
& A_m(2,3)=-A_0 - b_2 Q_0 -2 N_0 \frac{\gamma_{2,m}^2}{k_2^2},\ A_m(2,4)=A_m(2,3) \notag
\\
& A_m(2,5)= 2 N_0 \alpha_m \frac{\gamma_{3,m} }{k_3^2},\ A_m(2,6)=-A_m(2,5),\ A_m(2,8)=-{\rm j}\rho_i (1-\epsilon) \omega  \notag\\
& A_m(3,1)=-(1-\epsilon+b_1 \epsilon)\frac{\gamma_{1,m}\omega}{k_1^2},\ A_m(3,2)=-A_m(3,1) \notag
\\
& A_m(3,3)=-(1-\epsilon+b_2 \epsilon)\frac{\gamma_{2,m}\omega}{k_2^2},\ A_m(3,4)=-A_m(3,3) \notag
\\
& A_m(3,5)=(1-\epsilon+g \epsilon)\frac{\alpha_m \omega}{k_3^2},\ A_m(3,6)= A_m(3,5) \notag
\\
& A_m(3,8)={\rm j}\gamma_{i,m} \notag
\\
& A_m(4,1)=2 N_0\frac{\alpha_m \gamma_{1,m}}{k_1^2},\ A_m(4,2)=A_m(4,1),\ A_m(4,3)=2 N_0\frac{\alpha_m \gamma_{2,m}}{k_2^2},\ A_m(4,4)=A_m(4,3) \notag
\\
& A_m(4,5)= N_0\frac{-\alpha_m^2 + \gamma_{3,m}^2}{k_3^2},\ A_m(4,6)=A_m(4,5) \notag
\\
& A_m(5,1)=- (Q_0+b_1 R_0) {\rm e}^{-{\rm j} h_p \gamma_{1,m}},\ A_m(5,2)=- (Q_0+b_1 R_0) {\rm e}^{ {\rm j} h_p \gamma_{1,m}} \notag
\\
& A_m(5,3)=- (Q_0+b_2 R_0) {\rm e}^{-{\rm j} h_p \gamma_{2,m}},\ A_m(5,4)=- (Q_0+b_2 R_0) {\rm e}^{ {\rm j} h_p \gamma_{2,m}} \notag
\\
& A_m(5,9)=-{\rm j}\omega\epsilon \rho_a {\rm e}^{-{\rm j} h_p \gamma_{a,m}},\ A_m(5,10)=-{\rm j}\omega\epsilon \rho_a {\rm e}^{ {\rm j} h_p \gamma_{a,m}} \notag
\\
& A_m(6,1)=- (A_0+b_1 Q_0 + 2 N_0 \frac{\gamma_{1,m}^2}{k_1^2}) {\rm e}^{-{\rm j} h_p \gamma_{1,m}},\ A_m(6,2)=- (A_0+b_1 Q_0 + 2 N_0 \frac{\gamma_{1,m}^2}{k_1^2}) {\rm e}^{ {\rm j} h_p \gamma_{1,m}} \notag
\\
& A_m(6,3)=- (A_0+b_2 Q_0 + 2 N_0 \frac{\gamma_{2,m}^2}{k_2^2}) {\rm e}^{-{\rm j} h_p \gamma_{2,m}},\ A_m(6,4)=- (A_0+b_2 Q_0 + 2 N_0 \frac{\gamma_{2,m}^2}{k_2^2}) {\rm e}^{ {\rm j} h_p \gamma_{2,m}} \notag
\\
& A_m(6,5)=2 N_0 \frac{\alpha_m \gamma_{3,m}}{k_3^2} {\rm e}^{-{\rm j} h_p \gamma_{3,m}},\ A_m(6,6)=-2 N_0 \frac{\alpha_m \gamma_{3,m}}{k_3^2} {\rm e}^{ {\rm j} h_p \gamma_{3,m}} \notag
\\
& A_m(6,9)=-{\rm j}\omega(1-\epsilon) \rho_a {\rm e}^{-{\rm j} h_p \gamma_{a,m}},\ A_m(6,10)=-{\rm j}\omega(1-\epsilon) \rho_a {\rm e}^{ {\rm j} h_p \gamma_{a,m}} \notag
\\
& A_m(7,1)=- (1 - \epsilon + b_1 \epsilon) \frac{\gamma_{1,m} \omega}{k_1^2} {\rm e}^{-{\rm j} h_p \gamma_{1,m}},\ A_m(6,2)= (1 - \epsilon + b_1 \epsilon) \frac{\gamma_{1,m} \omega}{k_1^2} {\rm e}^{ {\rm j} h_p \gamma_{1,m}} \notag
\\
& A_m(7,3)=- (1 - \epsilon + b_2 \epsilon) \frac{\gamma_{2,m} \omega}{k_2^2} {\rm e}^{-{\rm j} h_p \gamma_{2,m}},\ A_m(7,4)= (1 - \epsilon + b_2 \epsilon) \frac{\gamma_{2,m} \omega}{k_2^2} {\rm e}^{ {\rm j} h_p \gamma_{2,m}} \notag
\\
& A_m(7,5)=(1 - \epsilon + g \epsilon) \frac{\alpha_m \omega}{k_3^2} {\rm e}^{-{\rm j} h_p \gamma_{3,m}},\ A_m(7,6)=(1 - \epsilon + g \epsilon) \frac{\alpha_m \omega}{k_3^2} {\rm e}^{ {\rm j} h_p \gamma_{3,m}} \notag
\\
& A_m(7,9)=-{\rm j} \gamma_{a,m} {\rm e}^{-{\rm j} h_p \gamma_{a,m}},\ A_m(7,10)= {\rm j} \gamma_{a,m} {\rm e}^{ {\rm j} h_p \gamma_{a,m}} \notag
\\
& A_m(8,1)=2 N_0 \frac{\alpha_m \gamma_{1,m} }{k_1^2} {\rm e}^{-{\rm j} h_p \gamma_{1,m}},\ A_m(8,2)=-2 N_0 \frac{\alpha_m \gamma_{1,m} }{k_1^2} {\rm e}^{ {\rm j} h_p \gamma_{1,m}} \notag
\\
& A_m(8,3)=2 N_0 \frac{\alpha_m \gamma_{2,m} }{k_2^2} {\rm e}^{-{\rm j} h_p \gamma_{2,m}},\ A_m(8,4)=-2 N_0 \frac{\alpha_m \gamma_{2,m} }{k_2^2} {\rm e}^{ {\rm j} h_p \gamma_{2,m}} \notag
\\
& A_m(8,5)=N_0 \frac{-\alpha_m^2 + \gamma_{3,m}^2}{k_3^2} {\rm e}^{-{\rm j} h_p \gamma_{3,m}},\ A_m(8,6)=N_0 \frac{-\alpha_m^2 + \gamma_{3,m}^2}{k_3^2} {\rm e}^{ {\rm j} h_p \gamma_{3,m}} \notag
\\
& A_m(9,7)={\rm j} \omega,\ A_m(9,9)=-{\rm j} \gamma_{a,m} {\rm e}^{-{\rm j} L_1 \gamma_{a,m}},\ A_m(9,10)= {\rm j} \gamma_{a,m} {\rm e}^{ {\rm j} L_1 \gamma_{a,m}} \notag
\\
& A_m(10,7)=-D \alpha_m^4 +\rho_p h_1 \omega^2,\ A_m(10,9)= {\rm j} \omega \rho_a {\rm e}^{-{\rm j} L_2 \gamma_{a,m}} \notag
\\
& A_m(10,10)= {\rm j} \omega \rho_a {\rm e}^{ {\rm j} L_2 \gamma_{a,m}},\ A_m(10,11)= -{\rm j} \omega \rho_t {\rm e}^{-{\rm j} L_2 \gamma_{t,m}} \notag
\\
& A_m(11,7)={\rm j} \omega,\ A_m(11,11)= -{\rm j} \gamma_{t,m} {\rm e}^{-{\rm j} L_3 \gamma_{t,m}}
\end{eqnarray}
where $Q_0, R_0, A_0, N_0$, $k_1, k_2, k_3, b_1, b_2$, and $g$ are detailed in Ref. \cite{Bolton1996JSV,Qiao2019JSV}.

The non-zero elements of vector ${\bf p}$ are: $p(1)={\rm j}\omega \epsilon \rho_i,\ p(2)={\rm j}\omega (1-\epsilon) \rho_i,\ p(3)={\rm j} k_z$. 

The non-zero elements of matrix ${\bf B}_m$ in multiple simple resonator cases are provided here to illustrate the core ideas. Denoting the row and column block index of full block matrix $\tilde{\bf B}$ as $m, n$, if $N_s >1$ and $m \ne n$, the non-zero element is
\begin{eqnarray}
B_m(10, 7)=-\frac{1}{l_x}\left[ \sum_{i=1}^{N_s-1} \frac{k_i m_i \omega^2}{k_i - m_i \omega^2} {\rm e}^{{\rm j} 2\pi (n-m) \frac{-i a}{l_x}} + \frac{1}{2} \frac{k_0 m_0 \omega^2}{k_0 - m_0 \omega^2} + \frac{1}{2} \frac{k_{\tiny N_s} m_{\tiny N_s} \omega^2}{k_{\tiny N_s} - m_{\tiny N_s} \omega^2} {\rm e}^{{\rm j} 2\pi (n-m) \frac{-N_s a}{l_x}} \right] \notag
\end{eqnarray}
If $N_s >1$ and $m = n$, the non-zero element is
\begin{eqnarray}
B_m(10, 7)=-\frac{1}{l_x}\left[ \sum_{i=1}^{N_s-1} \frac{k_i m_i \omega^2}{k_i - m_i \omega^2} + \frac{1}{2} \frac{k_0 m_0 \omega^2}{k_0 - m_0 \omega^2} + \frac{1}{2} \frac{k_{\tiny N_s} m_{\tiny N_s} \omega^2}{k_{\tiny N_s} - m_{\tiny N_s} \omega^2} \right] \notag
\end{eqnarray}
If $N_s =1$, the non-zero element is
\begin{eqnarray}
B_m(10, 7)=-\frac{1}{l_x} \frac{k_1 m_1 \omega^2}{k_1 - m_1 \omega^2}  \notag
\end{eqnarray}

\bibliography{mybibfile}

\end{document}